\documentclass[twocolumn]{aastex631}
\usepackage{graphicx}

\newcommand\kms{\rm km\ s^{-1}}

\accepted{September 27, 2024}

\submitjournal{AAS}

\shorttitle{Kinematic Insights Into LBVs and sgB[e] Stars}
\shortauthors{Deman and Oey}

\begin{document}

\title{
Kinematic Insights Into Luminous Blue Variables and B[e] Supergiants}

\author{Julian A. Deman}
\affiliation{Department of Astronomy, 323 West Hall, University of Michigan \\
1085 S. University Avenue \\
Ann Arbor, MI 48109-1107, USA}

\author{M. S. Oey}
\affiliation{Department of Astronomy, 323 West Hall, University of Michigan \\
1085 S. University Avenue \\
Ann Arbor, MI 48109-1107, USA}

\correspondingauthor{M. S. Oey}

\begin{abstract}

Recent work suggests that
many
luminous blue variables (LBVs) and B[e] supergiants (sgB[e]) are isolated, 
implying that they may be products of massive binaries, kicked by partner supernovae (SNe).  
However, the evidence is somewhat complex and controversial.
To test this scenario, we measure the proper-motion velocities for  these objects in the LMC and SMC, using Gaia Data Release 3. Our LMC results show that the kinematics, luminosities, and IR properties point to LBVs and sgB[e] stars being distinct classes. We find that Class 1 LBVs, which have dusty nebulae,
and sgB[e] stars both show velocity distributions comparable to that of SMC field OBe stars, which are known to have experienced SN kicks.  The sgB[e] stars are faster, plausibly due to their lower average masses.
However, Class 2 LBVs, which are luminous objects without dusty nebulae,
show no signs of acceleration, 
therefore suggesting that they are single stars, pre-SN binaries, 
or perhaps binary mergers. The candidate LBV Class 3 stars, which are dominated by hot dust, are all confirmed sgB[e] stars; their luminosities and velocities show that they simply represent the most luminous and massive of the sgB[e] class. There are very few SMC objects, but
the sgB[e] stars are faster than their LMC counterparts, which may be consistent with expectations that lower-metallicity binaries are tighter, causing faster ejections. We also examine the distinct class of dust-free, weak-lined sgB[e] stars, finding that the SMC objects have the fastest velocities of the entire sample. 

\end{abstract}


\section{Introduction} \label{sec:intro}
Luminous blue variables (LBVs) are an enigmatic class of hot, evolved, massive stars with a range of unusual properties that distinguish them from other supergiants and evolved massive stars. These include variability in photometric brightness and color, accompanied by spectral changes which can be quite varied in character. S Doradus variability, named after the eponymous LMC LBV, refers to variation in effective temperature and luminosity over timescales of several years to decades, while ``giant eruptions" refer to very large increases in brightness
up to several magnitudes over a period of months \citep{Humphreys1994,Koter1996,Weis2020}. 
LBVs are recognized by their S Dor variability, which is their defining parameter
as a distinct group;
but due to the infrequency of their enhanced mass loss state, they are hard to identify. In their quiescent state their spectra
can at least temporarily resemble those of other hot, massive stars such as Of/WN stars \citep{Weis2020}. 

LBV spectra are also often similar to those of B[e] supergiants (sgB[e]) \citep{Weis2020}, another class of hot, evolved stars with emission lines
and in particular, forbidden emission lines, 
in their spectra, indicating cooling and expanding circumstellar material \citep{Kraus2019}, with observations giving credence to the existence of surrounding dusty disks. There may be a link between these two groups of stars due to their similar properties and an overlap in their luminosities \citep[e.g.,][]{Kraus2019}.  
The relationship between LBVs and sgB[e] stars is unclear, but evidence is growing that they are distinct classes, based on their dust properties,
luminosity range, distribution on the H-R diagram,
and other factors \citep[e.g.,][]{Kraus2019, Humphreys2017a}.

Questions about the evolutionary status of LBVs and sgB[e] stars,
and their relationship,
have been debated for decades. Traditional, single-star
models of massive star evolution treat the LBV as a transitional state between main sequence O-type stars and stripped-envelope Wolf-Rayet (WR) stars, with the LBV corresponding to a phase when the stellar envelope is ejected, perhaps due to internal instabilities \citep[e.g.,][]{Humphreys1994, Langer1994, Maeder1994}.
This is referred to as the Conti scenario \citep{Conti1975, Conti1979}. 
This is a 
primary scenario for how WR stars can form from massive, young O-type stars, since a temporary LBV phase can permit a period of elevated mass-loss that is estimated to be enough to remove the envelope of the star, and as winds from O-type stars are recognized to be much weaker than originally thought
due to clumping in hot star winds,
with mass-loss rates consequently lowered \citep[e.g.,][]{Bouret2005, Smith2006}. 

However, \citet{Smith2015} find that LBVs and sgB[e] stars are isolated and tend to avoid massive star clusters, using O stars as a proxy for the location of clusters. 
More than half of the LBVs in their sample are not in clusters or OB associations, and a number of them are hundreds of parsecs from the nearest O-type star. They make the critical observation that LBVs appear to be even more isolated from O-type stars than are WR stars, therefore implying that LBVs cannot represent an intermediate evolutionary phase between these classes. They also find that isolation of sgB[e] stars from the nearest O-type stars is even greater than for LBVs, being comparable to that of red supergiants (RSGs).  This leads Smith \& Tombleson to suggest that sgB[e] stars are lower-mass analogs of LBVs, and that both LBVs and sgB[e] stars instead form in binaries, and are kicked after a supernova event. 

\citet{Humphreys2016} contest the findings by \citet{Smith2015},
revising their LBV sample and 
dividing it into ``classical LBVs" with masses $> 50\ M_\odot$, and ``less luminous LBVs" with masses between 25 and 40 $M_\odot$. They
demonstrate
that only classical LBVs should be associated with O-type stars, and find that 
the classical LBVs have a spatial distribution similar to that of late O-type stars and WR stars, while the less luminous LBVs have a distribution similar to RSGs.  They conclude that 
there is therefore no evidence that
classical LBVs are kicked from binaries. 

There has been continued controversy regarding the spatial distribution of LBVs, centered on both methodology and the LBV sample \citep{Nathan2016, Aghakhanloo2017, Aadland2018, Smith2019}.
On the other hand, there is more agreement that sgB[e] stars are deep field objects \citep{Smith2015, Humphreys2017a}, therefore supporting a binary origin scenario for this stellar class.  

The stellar kinematics offer a way to clarify the observational results regarding the spatial distribution of LBVs and sgB[e] stars.  \citet{Aghakhanloo2022} examine the radial velocities of LBVs in the LMC, and find  that over 33\% of LBVs have radial velocities 
exceeding 25 km/s.
In this work, we further examine whether LBVs and sgB[e] stars are ejected from clusters by measuring their peculiar proper motion velocities using data from the Gaia mission \citep{GaiaCollab2022}.

\section{samples and techniques}

The LMC and SMC offer a unique opportunity to study the proper motions of complete samples of these stars. The most comprehensive catalog of LBVs to date is the census for these galaxies by \citet{Agliozzo2021}, which is updated from that of \citet{Richardson2018}.  This compilation includes 9 LBVs in the LMC and 2 in the SMC; thus, the LBVs are statistically a rare class. \citet{Agliozzo2021} also list 
candidate LBV (LBVc) stars, which are spectroscopically similar to LBVs. Their magnitudes, colors, and masses overlap with those of LBVs, but they lack strong variable activity. Thus, LBVc stars may be dormant LBVs \citep[e.g.,][]{Smith2015,Solovyeva2020}. 
One such star that was previously considered a candidate LBV, HD 269582, was subsequently confirmed as a true LBV by \citet{Walborn2017}. 
There are 9 LBVc stars known in the LMC. \citet{Agliozzo2021} 
list two stars as LBVc in the SMC, but these are identified as sgB[e] stars by \citet{Kraus2019}, and we adopt the latter's classifications to define our sgB[e] sample.

SgB[e] stars are also rare, although not quite as rare as LBV/LBVc stars. The most comprehensive catalog of sgB[e] stars was compiled by \citet{Kraus2019}. This census identifies 13 confirmed sgB[e] stars and two candidate sgB[e] stars (hereafter sgB[e]c stars) in the LMC; and 5 confirmed sgB[e] stars and 1 sgB[e]c in the SMC. For both LMC and SMC, there are additionally a few related objects that are sometimes identified as sgB[e] stars, but which may be fundamentally different classes of objects, including the dust-poor,
weak-lined sgB[e] stars (hereafter sgB[e]-WL)
found by \citet{Graus2012}. Two more potential sgB[e] and LBVc stars were not included in the Kraus catalog: the star AzV 261 in the SMC \citep{Kalari2018}, and [BE74] 328 in the LMC \citep{Jones2015}, which we classify as sgB[e]c stars in this work, since they are not mentioned by either \citet{Kraus2019} or \citet{Agliozzo2021}. This leaves two sgB[e]c stars in each galaxy. 

The LMC LBVc star HD 269227 was not included in our sample due to a lack of reliable Gaia data;
it has a positive parallax and very large 
Gaia ``renormalized unit weight error" (RUWE), 
indicating poor data quality.
Thus, we have a total of 9 LBVs, 9 LBVc stars, 13 sgB[e] stars, 2 sgB[e]c stars, and 2 sgB[e]-WL stars for our LMC sample; and 2 LBVs, 0 LBVc stars, 5 sgB[e] stars, 2 sgB[e]c stars, and 3 sgB[e]-WL stars for our SMC sample. 

These designations include identifying R81, R99, and HD 5980 as candidate objects; their status as LBVs has been substantially questioned by \citet[e.g.,][]{Humphreys2016}. Both R81 and R99 appear in several other texts concerning LBVs, including \citet{Smith2015} and \citet{Aghakhanloo2022} without any concern for their status as LBVs. As such, we will continue to follow the classification given by \citet{Agliozzo2021}.
However, we caution that these stars may be fundamentally different objects and they are individually identified in some of our analysis below. 
Our adopted samples of LBV, LBVc, sgB[e], and sgB[e]c stars are given in Table~\ref{tab:samplelmc} for the LMC and Table~\ref{tab:samplesmc} for the SMC.  

We measure the residual proper motion velocities of our sample target stars using Gaia DR3 data \citep{GaiaCollab2022}, following the general method described by \citet{Oey2018} and \citet{Phillips2024}. For each target star, the local field velocity is obtained from the stars in the Gaia DR3 catalog within a $3\arcmin$ radius that have $G < 18$, excluding stars with incomplete basic data. We examine the RA and decl. velocities independently, discarding stars whose velocity or velocity error have values $> 3\sigma$ from the median for each component. The target star peculiar velocity is then calculated as the difference from the local systemic RA and decl. proper motion velocities. 

Tables~\ref{tab:vel_lmc} and \ref{tab:vel_smc} give the target star and local systemic velocities in the LMC and SMC, respectively. 
In these tables, columns 1 and 2 give the object name and stellar class, respectively, and column 3 lists the number of field stars used to determine the residual peculiar velocity.  Columns 4 and 5 give the target star R.A. and decl. 
velocity components of the
Gaia proper motions, and columns 6 and 7 give the R.A. and decl. 
velocity components of the
field systemic velocity, respectively, for each target; column 8 lists the target star's transverse peculiar velocity $v_\perp$.  Column 9 provides the stellar luminosity from the literature, which is a rough proxy for the stellar mass.
The quoted errors for the target stars are the Gaia measurement errors, and those for the local field are the standard error of the median. The total velocity errors are calculated by combining these errors in quadrature.
Tables~\ref{tab:LMCmedians} 
summarizes the characteristic $v_\perp$ for the different stellar classes in the LMC and SMC, including values calculated by weighting by the inverse of the measurement errors.  The table also lists the median of the errors on the $v_\perp$ measurements for each class. 

Since the earlier paper by \citet{Phillips2024} used slightly different definitions of field star samples, we also compared with residual velocities obtained by selecting the local field stars within radii of 3$\arcmin$ versus  
$5\arcmin$, $G< 17$ versus $G< 18$,
and using the Gaia DR3 catalog for field star selection versus the catalogs of \citet{Zaritsky2002} and \citet{Zaritsky2004}. Additionally, we compared the results using the outlier rejection method from \citet{Phillips2024} and our $3\sigma$ rejection method. None of these variations generate significantly different results within the errors. 

Given the much smaller samples of LBV and sgB[e] stars in the SMC, most of our findings focus on results for the LMC.  However, the SMC stars hint at possible significant differences, as discussed below.

\begin{deluxetable}{lllc}
\tablecaption{LMC Target Stars}\label{tab:samplelmc}
\tablehead{
\colhead{Object} & \colhead{Alternate Identifiers} & \colhead{Type\tablenotemark{a}} & \colhead{Class}\tablenotemark{a}} 
\startdata
HD 269006 & R 71 & LBV & 1b \\
HD 269216 &	SK $-69$ 75 & LBV & 1b \\
S Doradus & HD 35343 & LBV & 1a \\
HD 269582 &	SK $-69$ 142a & LBV & 2 \\
HD 269662 &	R 110 & LBV & 1b \\
HD 269700 &	R 116 & LBV & 2 \\
HD 269858 & R 127 & LBV & 1a \\
R 143 & CPD $-69$ 463 & LBV & 1b \\
HD 269321 & R 85 & LBV & \nodata \\
HD 268939 & R 74 & LBVc & 2 \\
LHA 120-S 18 & SK $-68$ 42 & LBVc & 1a \\
HD 269050 & R 78 & LBVc & 1b \\
HD 269128 &	R 81 & LBVc & 1a \\
HD 269445 & R 99 & LBVc & 2 \\
HD 269687 & LHA 120-S 119 & LBVc & 1a \\
HD 37836 & R 123 & LBVc & 2 \\
SK $-69$ 279 & {[BE74]} 619 & LBVc & 1a \\
LHA 120-S 61 & AL 418 & LBVc & 1a \\
\hline
HD 268835 & R 66 & sgB[e] & 3 \\
HD 34664 & MWC 105 & sgB[e] & 3 \\
HD 37974 & R 126 & sgB[e] & 3 \\
HD 38489 & SK $-69$ 259 & sgB[e] & 3 \\
LHA 120-S 12 & SK $-67$ 23 & sgB[e] & \nodata \\
LHA 120-S 35 & SK $-66$ 97 & sgB[e] & \nodata \\
LHA 120-S 59 & [BE74] 475 & sgB[e] & \nodata \\
LHA 120-S 89 & HD 269217 & sgB[e] & \nodata \\
LHA 120-S 93 & SK $-68$ 66 & sgB[e] & \nodata \\
LHA 120-S 111 &	HD 269599 & sgB[e] & \nodata \\
LHA 120-S 137 & {[BE74]} 621 & sgB[e] & \nodata \\
LHA 120-S 165 & {[BE74]} 587 & sgB[e] & \nodata \\
ARDB 54	& \nodata & sgB[e] & \nodata \\
VFTS 822 & \nodata & sgB[e]c & \nodata \\
{[BE74]} 328 & SK $-69$ 169 & sgB[e]c\tablenotemark{b} & \nodata \\
\hline
VFTS 698 & AL 377 & sgB[e]-WL & \nodata \\
{[L72]} LH 85-10 & [BE74] 388 & sgB[e]-WL\tablenotemark{c} & \nodata \\
\enddata
\tablenotetext{a}{All LBVs and LBVc types and Classes are from \citet{Agliozzo2021}, and all sgB[e] stars and sgB[e]c are from \citet{Kraus2019}, with the exception of {[BE74]} 328.
Class types for sgB[e]-WL stars are from 
\citet{Graus2012}, with the exception of [L72] LH 85-10.
}
\tablenotetext{b}{From \citet{Jones2015}}
\tablenotetext{c}{This work}
\end{deluxetable}

\begin{deluxetable}{lllc}
\tablecaption{SMC Target Stars} \label{tab:samplesmc}
\tablehead{
\colhead{Object} & \colhead{Alternate Identifiers} & \colhead{Type\tablenotemark{a}} & \colhead{Class\tablenotemark{a}} }
\startdata
HD 5980	& R 14 & LBV & 2 \\
HD 6884 & R 40 & LBV & 1b \\
\hline
LHA 115-S 6	& R 4 & sgB[e] & 3 \\
LHA 115-S 18 & AzV 154 & sgB[e] & 3 \\
LHA 115-S 23 & AzV 172 & sgB[e] & \nodata \\
LHA 115-S 65 & R 50 & sgB[e] & \nodata \\
{[MA93]} 1116 & \nodata & sgB[e] & \nodata \\
LHA 115-S 38 & [MA93] 1405 & sgB[e]c & \nodata \\
AzV 261 & [MA93] 1235 & sgB[e]c\tablenotemark{b} & \nodata \\
\hline
LHA 115-S 29 & [MA93] 1149 & sgB[e]-WL & \nodata \\
LHA 115-S 46 & [MA93] 1552 & sgB[e]-WL & \nodata \\
LHA 115-S 62 & R 48 & sgB[e]-WL & \nodata \\
\enddata
\tablenotetext{a}{All LBVs and their Classes are from \citet{Agliozzo2021}, and all sgB[e] stars and sgB[e]c stars are from \citet{Kraus2019}, with the exception of AzV 261.
Class types for sgB[e]-WL stars are from 
\citet{Graus2012}.}
\tablenotetext{b}{From \citet{Kalari2018}}
\end{deluxetable}

\begin{deluxetable*}{llccccccc}
\tablecaption{LMC Proper Motion Measurements} \label{tab:vel_lmc}
\tablehead{
\colhead{Object} & \colhead{Type} & \colhead{Number of} & \colhead{ {$v_\alpha$}(Target)} & \colhead{{$v_\delta$}(Target)} & \colhead{ {$v_\alpha$}(Field)} & \colhead{{$v_\delta$}(Field)} & \colhead{$v_{\perp}$} & \colhead{Luminosity\tablenotemark{a}} \\
\colhead{} & \colhead{} & \colhead{field stars} & \colhead{$\kms$} & \colhead{$\kms$} & \colhead{$\kms$} & \colhead{$\kms$} & \colhead{$\kms$} & \colhead{$\log(L/L_\sun)$}}
\startdata
HD 269006 & LBV & 71 & 428.4 $\pm{4.6}$ & -28.1 $\pm{6.0}$ & 479.1 $\pm{4.1}$ & -6.09 $\pm{5.1}$ & 55.2 $\pm{10.0}$ & 5.78 \\
HD 269216 & LBV & 558 & 411.9 $\pm{4.0}$ & 122.1 $\pm{4.3}$ & 413.8 $\pm{1.6}$ & 116.3 $\pm{3.2}$ & 22.8 $\pm{7.5}$ & 5.44 \\
S Doradus & LBV & 733 & 420.4 $\pm{4.2}$ & 88.0 $\pm{5.3}$ & 421.5 $\pm{2.4}$ & 68.4 $\pm{1.2}$ & 19.7 $\pm{7.2}$ & 6.3 \\
HD 269582 & LBV & 552 & 395.8 $\pm{5.9}$ & 152.9 $\pm{5.5}$ & 394.0 $\pm{2.0}$ & 96.3 $\pm{5.0}$ & 56.6 $\pm{9.7}$ & 5.68 \\
HD 269662 & LBV & 333 & 401.1 $\pm{5.9}$ & 119.3 $\pm{6.4}$ & 397.6 $\pm{2.4}$ & 114.9 $\pm{2.8}$ & 5.6 $\pm{9.4}$ & 5.36 \\
HD 269700 & LBV & 167 & 381.1 $\pm{5.2}$ & 117.1 $\pm{6.2}$ & 385.1 $\pm{2.7}$ & 125.8 $\pm{0.3}$ & 9.6 $\pm{8.5}$ & 5.92 \\
HD 269858 & LBV & 558 & 411.9 $\pm{4.0}$ & 122.1 $\pm{4.3}$ & 413.8 $\pm{1.6}$ & 116.3 $\pm{3.2}$ & 6.1 $\pm{6.9}$ & 6 \\
R 143 & LBV & 417 & 412.4 $\pm{4.4}$ & 141.0 $\pm{4.3}$ & 396.1 $\pm{2.3}$ & 145.0 $\pm{3.5}$ & 16.7 $\pm{7.5}$ & 5.32 \\
HD 269321 & LBV & 666 & 420.4 $\pm{5.2}$ & 96.0 $\pm{5.2}$ & 428.9 $\pm{2.5}$ & 68.8 $\pm{1.1}$ & 28.5 $\pm{7.9}$ & 5.67 \\
HD 268939 & LBVc & 197 & 389.0 $\pm{4.3}$ & 1.1 $\pm{4.3}$ & 396.2 $\pm{2.4}$ & -5.7 $\pm{0.2}$ & 9.9 $\pm{6.5}$ & 5.4 \\
LHA 120-S 18 & LBVc & 195 & 419.4 $\pm{5.8}$ & 20.0 $\pm{6.2}$ & 420.2 $\pm{2.7}$ & 16.7 $\pm{0.1}$ & 3.4 $\pm{8.9}$ & 5.58 \\
HD 269050 & LBVc & 357 & 437.6 $\pm{6.0}$ & 37.8 $\pm{6.9}$ & 433.6 $\pm{2.2}$ & 26.0 $\pm{3.1}$ & 12.5 $\pm{10.0}$ & 5.78 \\
HD 269128 & LBVc & 472 & 449.4 $\pm{5.2}$ & 44.0 $\pm{5.8}$ & 423.7 $\pm{2.3}$ & 34.3 $\pm{3.3}$ & 27.4 $\pm{8.8}$ & 6 \\
HD 269445 & LBVc & 144 & 391.4 $\pm{4.0}$ & 67.1 $\pm{4.1}$ & 389.5 $\pm{2.8}$ & 69.8 $\pm{2.5}$ & 3.3 $\pm{6.9}$ & 6.26 \\
HD 269687 & LBVc & 312 & 435.4 $\pm{7.7}$ & 88.2 $\pm{7.7}$ & 395.5 $\pm{2.8}$ & 121.3 $\pm{1.1}$ & 51.8 $\pm{11.3}$ & 5.78 \\
HD 37836 & LBVc & 424 & 415.4 $\pm{4.1}$ & 117.8 $\pm{4.3}$ & 419.4 $\pm{2.1}$ & 124.7 $\pm{3.9}$ & 7.9 $\pm{7.4}$ & 6.5 \\
Sk $-69$ 279 & LBVc & 314 & 428.7 $\pm{5.5}$ & 184.6 $\pm{6.6}$ & 420.2 $\pm{2.0}$ & 151.3 $\pm{0.9}$ & 34.3 $\pm{8.9}$ & 5.6 \\
LHA 120-S 61 & LBVc & 132 & 448.9 $\pm{6.9}$ & 163.3 $\pm{6.9}$ & 366.0 $\pm{2.7}$ & 151.1 $\pm{0.9}$ & 83.8 $\pm{10.1}$ & 5.8 \\
\hline
HD 268835 & sgB[e] & 200 & 444.4 $\pm{4.3}$ & -15.8 $\pm{5.3}$ & 462.6 $\pm{2.8}$ & -19.0 $\pm{0.5}$ & 18.5 $\pm{7.4}$ & 5.47 \\
HD 34664 & sgB[e] & 133 & 395.4 $\pm{5.5}$ & 41.0 $\pm{5.5}$ & 384.2 $\pm{2.9}$ & 24.7 $\pm{3.6}$ & 19.8 $\pm{9.1}$ & 5.56 \\
HD 37974 & sgB[e] & 480 & 409.8 $\pm{3.7}$ & 126.6 $\pm{3.8}$ & 408.0 $\pm{1.8}$ & 120.8 $\pm{3.3}$ & 6.1 $\pm{6.5}$ & 6.15 \\
HD 38489 & sgB[e] & 434 & 426.2 $\pm{5.4}$ & 144.8 $\pm{6.0}$ & 412.9 $\pm{1.9}$ & 142.8 $\pm{2.0}$ & 13.5 $\pm{8.5}$ & 6.13 \\
LHA 120-S 12 & sgB[e] & 153 & 432.7 $\pm{4.0}$ & 0.4 $\pm{4.4}$ & 422.7 $\pm{2.6}$ & -23.2 $\pm{1.9}$ & 25.7 $\pm{6.7}$ & 5.41\tablenotemark{b} \\
LHA 120-S 35 & sgB[e] & 193 & 353.4 $\pm{4.8}$ & 138.5 $\pm{5.5}$ & 367.9 $\pm{2.0}$ & 107.2 $\pm{2.3}$ & 34.5 $\pm{7.9}$ & 5.21\tablenotemark{b} \\
LHA 120-S 59 & sgB[e] & 107 & 333.8 $\pm{4.5}$ & 175.8 $\pm{4.3}$ & 371.3 $\pm{3.3}$ & 154.3 $\pm{0.3}$ & 43.3 $\pm{7.1}$ & 4.01\tablenotemark{b} \\
LHA 120-S 89 & sgB[e] & 582 & 469.4 $\pm{3.8}$ & 57.9 $\pm{3.9}$ & 448.2 $\pm{2.5}$ & 55.2 $\pm{0.8}$ & 21.4 $\pm{6.0}$ & 5.44\tablenotemark{b} \\
LHA 120-S 93 & sgB[e] & 161 & 419.2 $\pm{4.6}$ & 78.4 $\pm{5.1}$ & 416.5 $\pm{3.2}$ & 49.3 $\pm{2.0}$ & 29.3 $\pm{7.8}$ & 4.61\tablenotemark{b} \\
LHA 120-S 111 & sgB[e] & 592 & 386.3 $\pm{4.4}$ & 133.1 $\pm{4.7}$ & 398.8 $\pm{1.8}$ & 102.0 $\pm{3.5}$ & 33.5 $\pm{7.5}$ & 6.06\tablenotemark{b} \\
LHA 120-S 137 & sgB[e] & 301 & 428.7 $\pm{3.9}$ & 120.2 $\pm{4.7}$ & 426.3 $\pm{2.3}$ & 147.6 $\pm{1.5}$ & 27.5 $\pm{6.7}$ & 4.24\tablenotemark{b} \\
LHA 120-S 165 & sgB[e] & 172 & 523.9 $\pm{4.9}$ & 88.5 $\pm{5.0}$ & 491.5 $\pm{3.0}$ & 92.8 $\pm{5.3}$ & 32.7 $\pm{9.3}$ & 5.0\tablenotemark{c}\\
ARDB 54 & sgB[e] & 189 & 472.0 $\pm{3.2}$ & -30.4 $\pm{4.0}$ & 460.7 $\pm{2.4}$ & -19.8 $\pm{0.5}$ & 15.5 $\pm{5.7}$ & 4.4\tablenotemark{b}\\
VFTS 822 & sgB[e]c & 126 & 367.2 $\pm{7.7}$ & 176.2 $\pm{7.9}$ & 407.9 $\pm{3.9}$ & 147.1 $\pm{0.3}$ & 50.0 $\pm{11.7}$ & 4.04\tablenotemark{d} \\
{[BE74]} 328 & sgB[e]c & 402 & 396.6 $\pm{6.0}$ & 113.7 $\pm{6.6}$ & 395.1 $\pm{2.1}$ & 112.6 $\pm{1.2}$ & 1.8 $\pm{9.2}$ & 3.79\tablenotemark{e} \\
\hline
VFTS 698 & sgB[e]-WL & 481 & 409.5 $\pm{6.6}$ & 162.7 $\pm{6.6}$ & 395.5 $\pm{2.4}$ & 148.3 $\pm{4.6}$ & 20.1 $\pm{10.7}$ & 5.6\tablenotemark{f} \\
{[L72]} LH 85-10 & sgB[e]-WL & 318 & 370.4 $\pm{5.5}$ & 147.2 $\pm{5.4}$ & 381.4 $\pm{1.9}$ & 134.6 $\pm{3.1}$ & 16.7 $\pm{8.6}$ & 5.46\tablenotemark{g} \\
\enddata
\tablenotetext{a}{From \citet{Agliozzo2021} unless otherwise noted.}
\tablenotetext{b}{From \citet{Kraus2019}.}
\tablenotetext{c}{From \citet{Oksala2012}.}
\tablenotetext{d}{From \citet{Kalari2014}.}
\tablenotetext{e}{From \citet{Jones2015}.}
\tablenotetext{f}{From \citet{Dunstall2012}.}
\tablenotetext{g}{Calculated from bolometric magnitude of \citet{Massey2000}.}
\end{deluxetable*}

\centerwidetable
\begin{deluxetable*}{llccccccc}
\tablecaption{SMC Proper Motion Measurements} \label{tab:vel_smc}
\tablehead{
\colhead{Object} & \colhead{Type} & \colhead{Number of} & \colhead{ {$v_\alpha$}(Target)} & \colhead{{$v_\delta$}(Target)} & \colhead{ {$v_\alpha$}(Field)} & \colhead{{$v_\delta$}(Field)} & \colhead{$v_{\perp}$} & \colhead{Luminosity} \\
\colhead{} & \colhead{} & \colhead{field stars} & \colhead{$\kms$} & \colhead{$\kms$} & \colhead{$\kms$} & \colhead{$\kms$} & \colhead{$\kms$} & \colhead{$\log(L/L_\sun)$}}
\startdata
HD 5980 & LBV & 410 & 265.0 $\pm{10.2}$ & -392.8 $\pm{9.1}$ & 216.9 $\pm{2.7}$ & -365.6 $\pm{2.1}$ & 55.3 $\pm{14.1}$ & 6.5\tablenotemark{a} \\
HD 6884 & LBV & 229 & 255.2 $\pm{3.8}$ & -369.7 $\pm{3.9}$ & 248.6 $\pm{3.4}$ & -360.1 $\pm{1.2}$ & 11.6 $\pm{6.5}$ & 5.65\tablenotemark{a} \\
\hline
LHA 115-S 6 & sgB[e] & 435 & 224.3 $\pm{5.9}$ & -393.7 $\pm{4.8}$ & 188.5 $\pm{2.3}$ & -371.4 $\pm{2.0}$ & 42.4 $\pm{8.2}$ & 5.02\tablenotemark{b} \\
LHA 115-S 18 & sgB[e] & 394 & 221.9 $\pm{6.2}$ & -393.6 $\pm{5.6}$ & 199.5 $\pm{2.8}$ & -365.5 $\pm{0.6}$ & 35.9 $\pm{8.8}$ & 5.60\tablenotemark{b} \\
LHA 115-S 23 & sgB[e] & 233 & 200.7 $\pm{4.2}$ & -362.9 $\pm{4.0}$ & 209.4 $\pm{3.3}$ & -364.0 $\pm{0.9}$ & 8.8 $\pm{6.8}$ & 4.31\tablenotemark{b} \\
LHA 115-S 65 & sgB[e] & 86 & 370.1 $\pm{5.7}$ & -360.6 $\pm{5.5}$ & 340.5 $\pm{10.1}$ & -342.1 $\pm{3.4}$ & 35.0 $\pm{13.3}$ & 5.65\tablenotemark{b} \\
{[MA93]} 1116 & sgB[e] & 431 & 234.3 $\pm{18.4}$ & -347.7 $\pm{15.8}$ & 213.1 $\pm{2.6}$ & -366.4 $\pm{2.4}$ & 28.3 $\pm{24.5}$ & 4.4\tablenotemark{c} \\
LHA 115-S 38 & sgB[e]c & 265 & 217.0 $\pm{4.8}$ & -377.9 $\pm{4.5}$ & 232.9 $\pm{2.9}$ & -365.7 $\pm{3.4}$ & 20.0 $\pm{8.0}$ & 4.10\tablenotemark{b} \\
AzV 261 & sgB[e]c & 225 & 229.5 $\pm{5.5}$ & -370.1 $\pm{4.9}$ & 224.9 $\pm{3.0}$ & -364.0 $\pm{2.2}$ & 7.7 $\pm{8.3}$ & 4.9\tablenotemark{d} \\
\hline
LHA 115-S 29 & sgB[e]-WL & 188 & 271.6 $\pm{5.8}$ & -383.4 $\pm{4.7}$ & 223.0 $\pm{3.5}$ & -360.5 $\pm{0.7}$ & 53.7 $\pm{8.3}$ & 4.79\tablenotemark{e} \\
LHA 115-S 46 & sgB[e]-WL & 110 & 268.9 $\pm{5.4}$ & -331.5 $\pm{6.6}$ & 252.7 $\pm{4.2}$ & -363.7 $\pm{1.9}$ & 36.0 $\pm{9.7}$ & 4.61\tablenotemark{e} \\
LHA 115-S 62 & sgB[e]-WL & 114 & 264.9 $\pm{5.4}$ & -322.3 $\pm{4.8}$ & 307.0 $\pm{8.8}$ & -354.1 $\pm{1.0}$ & 52.8 $\pm{11.4}$ & 4.36\tablenotemark{e} \\
\enddata
\tablenotetext{a}{From \citet{Agliozzo2021}.}
\tablenotetext{b}{From \citet{Kraus2019}.}
\tablenotetext{c}{From \citet{Wisniewski2007}.}
\tablenotetext{d}{From \citet{Kalari2018}.}
\tablenotetext{e}{From \citet{Graus2012}.}
\end{deluxetable*}

\section{LBV velocity distributions}

\begin{figure}
\plotone{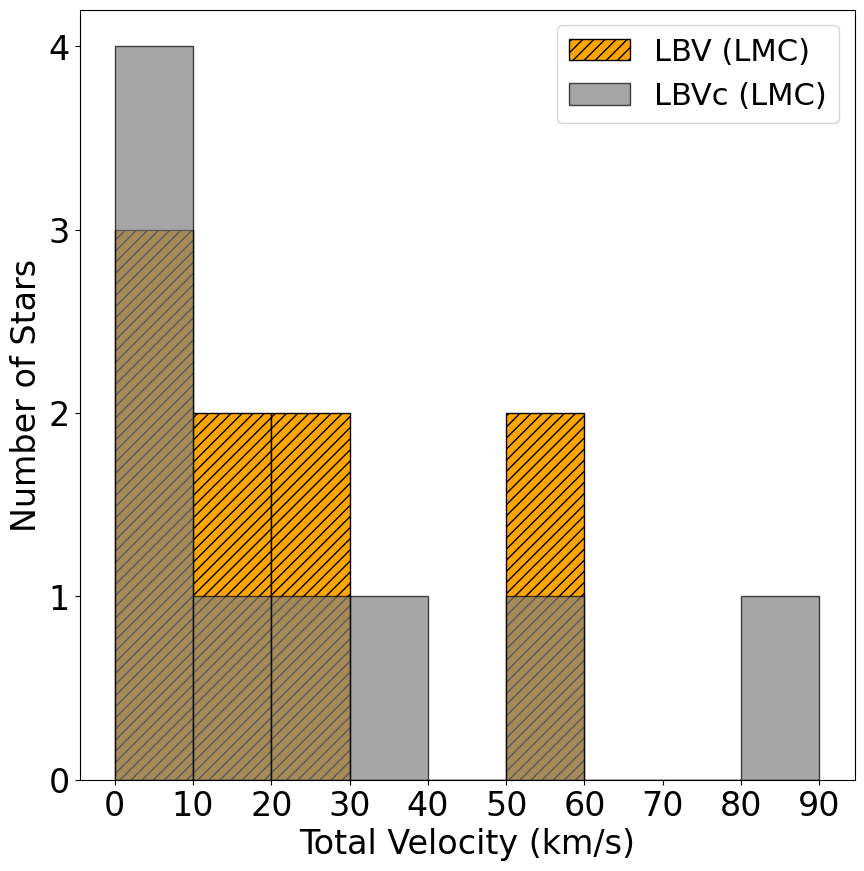}
\caption{Histogram comparing velocities of LBVs and LBVc stars in the LMC.}
\label{fig:LBVLBVc}
\end{figure}

Focusing first on the LBV stars, Figure~\ref{fig:LBVLBVc} compares the velocity distributions of LBV and LBVc stars in the LMC.  The two distributions look similar, and a 
two-sided Kolmogorov-Smirnov (K-S) test 
carried out with the Scipy package gives a $p$-value of 0.99, consistent with the possibility
that LBVc stars are quiescent LBVs. 
We caution that our sample sizes are small, with 9 stars in each of the two groups, hence the results cannot be regarded as conclusive.
But based on their kinematics, the LBV and LBVc stars do not show significant differences, and we therefore combine them into a single, larger sample for the purposes of this study.  This does not mean that significant differences may not exist between them.  In this section, we examine some subsets of this combined sample, and we also provide data for individual stars in Tables~\ref{tab:vel_lmc} and \ref{tab:vel_smc} so that studies can be done using different subsets and classifications for this group.
In what follows, our sample of LBVs thus includes the LBVc stars.

\begin{figure}
\plotone{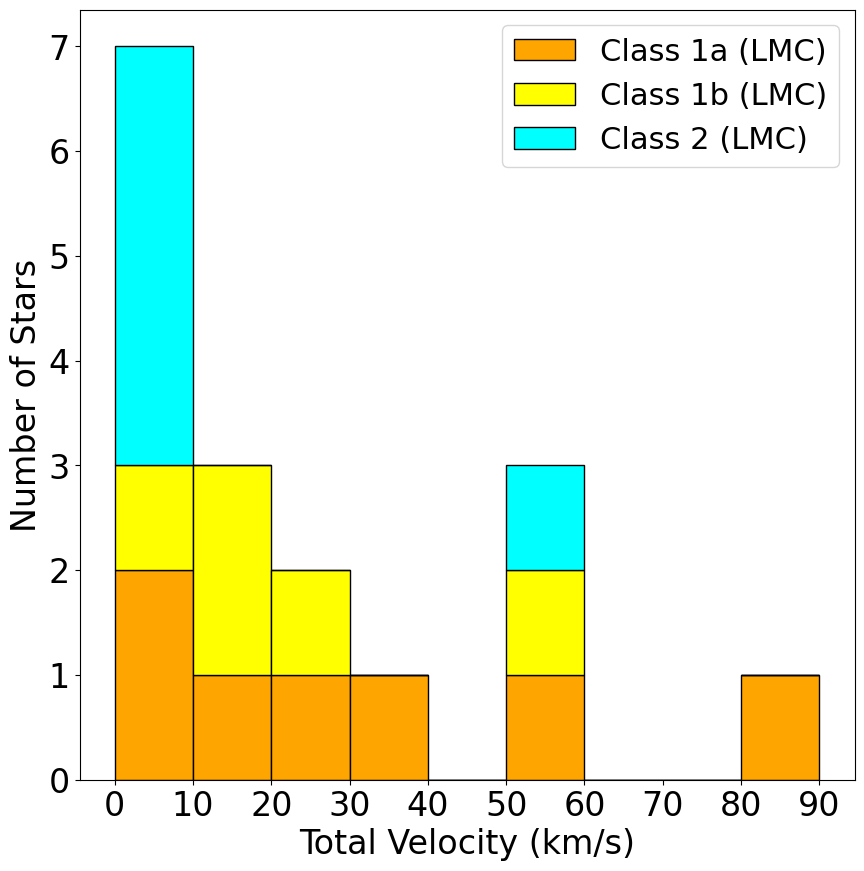}
\caption{Total velocity distribution of LBV and LBVc stars in the LMC, showing the contributions of objects in Classes 1a, 1b, and 2 indicated by stacked orange, yellow, and cyan bars respectively.}
\label{fig:LBV_SED}
\end{figure} 

\begin{deluxetable*}{lcccccc}\label{tab:LMCmedians}
\tablecaption{Characteristic Transverse Peculiar Velocities by Stellar Class}
\tablehead{
\colhead{Class} & \colhead{ $N$} & \colhead{Median} & \colhead{Median Error\tablenotemark{a}} & \colhead{Weighted Median} & \colhead{Weighted Mean} & \colhead{Standard Deviation} \\
\colhead{} & \colhead{} & \colhead{$\kms$} & \colhead{$\kms$} & \colhead{$\kms$} & \colhead{$\kms$} & \colhead{$\kms$}}
\startdata
LMC - All & 35 & 20.1 & 7.9 & 19.8 & 23.4 & 18.3 \\
LMC - LBV & 9 & 19.7 & 7.9 & 19.7 & 23.2 & 19.4 \\
LMC - LBVc & 9 & 12.5 & 8.9 & 9.9 & 23.1 & 27.1 \\
LMC - LBV + LBVc\tablenotemark{b} & 18 & 18.2 & 8.7 & 16.8 & 23.2 & 22.9 \\
LMC - Class 1a & 7 & 27.4 & 8.9 & 27.4 & 29.4 & 28.1 \\
LMC - Class 1b & 5 & 16.8 & 9.4 & 16.8 & 22.1 & 19.3 \\
LMC - Class 1 & 12 & 21.3 & 8.9 & 19.7 & 26.4 & 24.3 \\
LMC - Class 2 & 5 & 9.6 & 7.4 & 9.6 & 15.3 & 22.0 \\
LMC - Class 3 & 4 & 16.0 & 8.0 & 13.5 & 13.9 & 6.2 \\
LMC - sgB[e] non-Class 3  & 11 & 29.3 & 7.5 & 27.5 & 27.8 & 13.0 \\
LMC - sgB[e]\tablenotemark{c} & 15 & 25.7 & 7.5 & 25.7 & 24.9 & 13.1 \\
LMC - sgB[e]-WL & 2 & 18.4 & 9.6 & 16.7 & 18.2 & 2.4 \\
\hline
SMC - All & 12 & 35.4 & 8.6 & 35.0 & 29.9 & 17.3 \\
SMC - LBV\tablenotemark{b} & 2 & 33.4 & 10.3 & 11.6 & 25.4 & 30.8 \\
SMC - Class 1 & 1 & 11.6 & 6.5 & 11.6 & 11.6 & \nodata \\
SMC - Class 2\tablenotemark{d} & 1 & 55.9 & 14.1 & 55.9 & 55.9 & \nodata \\
SMC - Class 3 & 2 & 39.0 & 8.5 & 42.2 & 39.1 & 4.5 \\
SMC - sgB[e] non-Class 3 & 5 & 20.0 & 8.3 & 8.8 & 16.7 & 12.0 \\
SMC - sgB[e]\tablenotemark{c} & 7 & 28.3 & 8.3 & 20.0 & 23.8 & 13.6 \\
SMC - sgB[e]-WL & 3 & 52.8 & 9.7 & 52.8 & 47.6 & 9.9 \\
\hline
SMC - Field OBe\tablenotemark{e} & 160 & 25.5 & 14.5 & 24.0 & 29.9 & 21.6 \\
\enddata
\tablenotemark{a}{Median of measurement errors on $v_\perp$}
\tablenotetext{b}{Excluding Class 3.}
\tablenotetext{c}{Including sgB[e]c and excluding sgB[e]-WL.}
\tablenotetext{d}{This object is HD 5980.}
\tablenotetext{e}{From \citet{Phillips2024}}
\end{deluxetable*}

Massive star kinematics are often linked to binary interaction history \citep[e.g.,][]{Renzo2019, DorigoJones2020}, which also often plays a role in whether and when a star ejects material.
\citet{Agliozzo2021} classified LBVs and LBVc stars according to infrared environment and circumstellar emission.  
Class 1 corresponds to sources with dusty nebulae peaking in the mid- to far-infrared, 
suggesting previous substantial stellar mass ejection.
The Class 2 sources 
lack dusty nebulae, and
are dominated by free-free emission from 1 to 24 $\mu$m; they are thus less likely to have experienced mass ejection of the type or scale of Class 1 objects. 
\citet{Agliozzo2021} further divided Class 1 into Class 1a objects which have free-free excess above the stellar photosphere, and Class 1b objects without free-free excess. 
\citet{Humphreys2017b} suggest that 
the definition of LBVs should include a lack of warm dust emission in the IR,
which thus includes the Class~2 objects. As shown below, {\it the Class 2 }
{\it objects indeed appear to be a distinct physical class.} 
About half of the Class~1 objects would be considered by \citet{Humphreys2016} to belong to the ``less luminous'' LBV-like objects;
in Section~\ref{sec:classical}, we show results for confirmed LBVs classified according to this scheme.

\citet{Agliozzo2021} also specify Class 3 as sources dominated by a hotter, dusty component at wavelengths $> 2\ \mu$m; however, these objects are all known to be identified as sgB[e] stars and are catalogued as such by \citet[][; Table~\ref{tab:samplelmc}]{Kraus2019}. 
In Section~\ref{sec:class3} below, we show that the properties of Class 3 stars are consistent with their being luminous sgB[e] stars, and
we therefore exclude them from our LBV sample.  Data for the characteristic transverse velocities of Classes 1, 2, and 3 are included in Table~\ref{tab:LMCmedians}.

\begin{figure}
\plotone{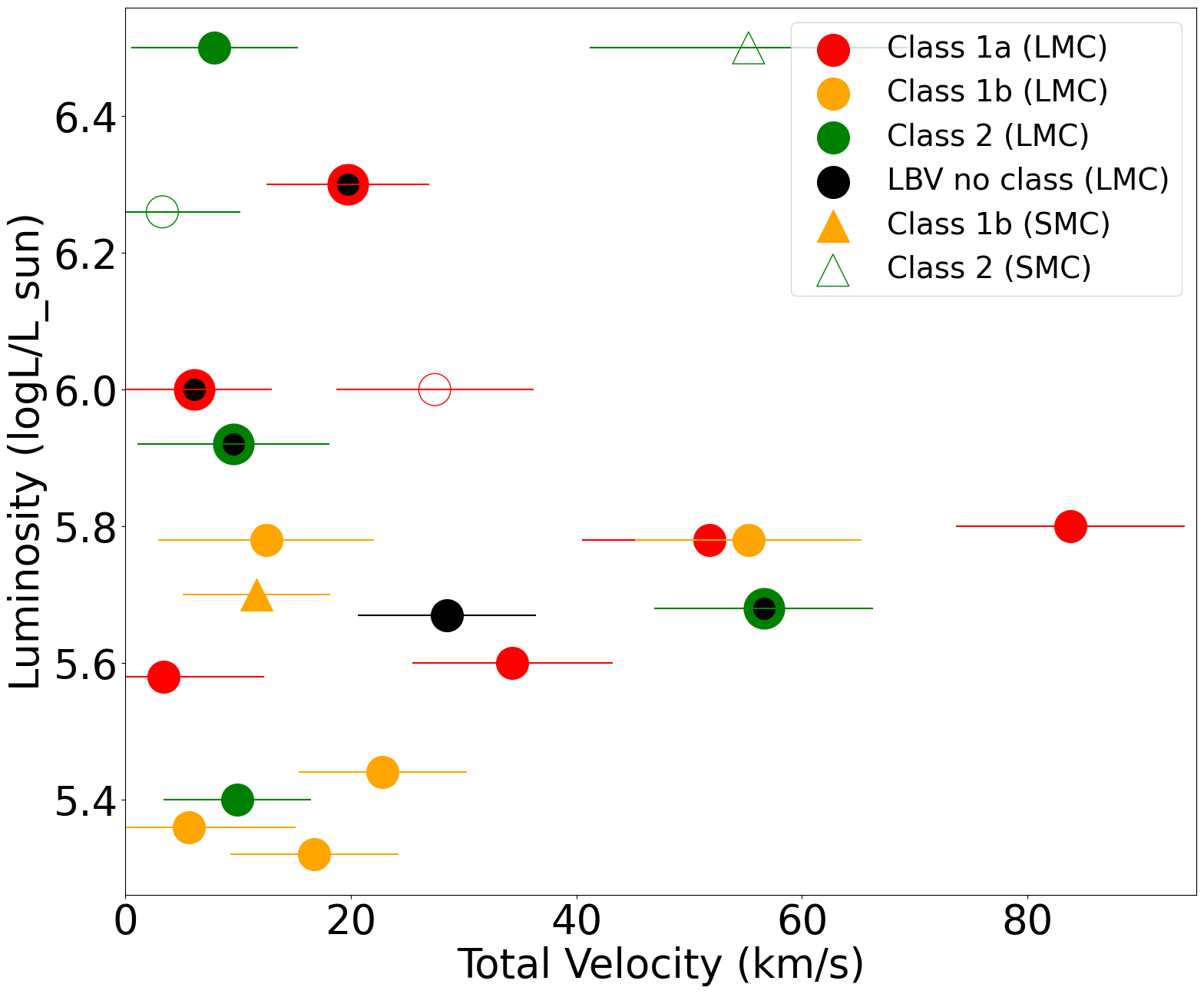}
\caption{
Luminosity vs $v_\perp$ for Classes 1a, 1b, and 2, 
color-coded as shown.  Circles and triangles correspond to LMC and SMC objects, respectively. 
The star R85 does not have an IR class and thus appears as a black circle. The SMC LBVc star HD 5980, which may not be a true LBV, is indicated with the open triangle, while the two open circles correspond to the stars R81 and R99, which also also have questionable LBV status  \citep{Humphreys2016}.
The four
stars considered to be ``classical" LBVs appear with black dots superposed on their symbols (see Section~\ref{sec:classical}).}
\label{fig:scatterLBVLMC}
\end{figure}

As seen in Table~\ref{tab:samplelmc}, the majority of LBV stars (12) belong to Class 1, and 5 belong to Class 2. Figure~\ref{fig:LBV_SED} shows the
contributions of Classes 1a, 1b, and 2 to the total velocity distribution of the LBV stars. We see that Classes 1a and 1b have similar distributions although Class 1a tends to have faster velocities than Class 1b,
with weighted median values of 27 $\kms$ vs 17 $\kms$, and weighted means
of 29 $\kms$ vs 22 $\kms$, respectively (Table~\ref{tab:LMCmedians}). However, the distributions are similar and dominated by small-number statistics.
Class 1a is notably more luminous than Class 1b, as can be seen in Table~\ref{tab:vel_lmc} and Figure~\ref{fig:scatterLBVLMC};
both groups show high velocities, consistent with a population dominated by ejections from the parent clusters.

On the other hand, Class 2 appears to be a distinct population, having velocities $<10\ \kms$, with the exception of HD 269582, whose velocity is 56.6 $\kms$. The weighted median velocity for Class 2 is 9.6 $\kms$, which is half the median value of 20 $\kms$ for all of Class 1 (Table~\ref{tab:LMCmedians}), and with the removal of HD 269582, the weighted median reduces further to 7.9 $\kms$. Class 2 stars also tend to be more luminous than either Class 1a or Class 1b (Table~\ref{tab:LMCluminosities} and Figure~\ref{fig:scatterLBVLMC}).
The low measured peculiar velocities are consistent with their not being accelerated by any ejection mechanism, although
they might also be merger products, which generally have much lower velocities.

\begin{deluxetable*}{lcccc}\label{tab:LMCluminosities}
\tablecaption{Characteristic Luminosities by Stellar Class}
\tablehead{
\colhead{Class} & \colhead{$N$} & \colhead{Mean} & \colhead{Median} & \colhead{Standard Deviation} \\
\colhead{} & \colhead{} & \colhead{$\log(L/L_\sun)$} & \colhead{$\log(L/L_\sun)$} & \colhead{$\log(L/L_\sun)$}}
\startdata
LMC - Class 1a & 7 & 5.93 & 5.80 & 0.25 \\
LMC - Class 1b & 5 & 5.58 & 5.44 & 0.23 \\
LMC - Class 1 & 12 & 5.81 & 5.79 & 0.32 \\
LMC - Class 2 & 5 & 6.12 & 5.92 & 0.44 \\
LMC - Class 3 & 4 & 5.93 & 5.85 & 0.36 \\
LMC - sgB[e] non-Class 3 & 11 & 5.41 & 4.51 & 0.76 \\
LMC - sgB[e] & 15 & 5.62 & 5.25 & 0.81 \\
LMC - sgB[e]-WL & 2 & 5.54 & 5.54 & 0.17 \\
\hline
SMC - LBV\tablenotemark{a} & 2 & 6.26 & 6.08 & 0.59 \\
SMC - Class 1 & 1 & 5.66 & 5.66 & \nodata \\
SMC - Class 2 & 1 & 6.50 & 6.50 & \nodata \\
SMC - Class 3 & 2 & 5.30 & 5.31 & 0.41 \\
SMC - sgB[e] non-Class 3 & 5 & 5.07 & 4.40 & 0.62 \\
SMC - All sgB[e]\tablenotemark{b} & 7 & 5.26 & 4.90 & 0.62 \\
SMC - sgB[e]-WL & 3 & 4.62 & 4.61 & 0.22 \\
\enddata
\tablenotetext{a}{Excluding Class 3.}
\tablenotetext{b}{Excluding sgB[e]-WL stars.}
\end{deluxetable*}

\subsection{LBVs as possible binary products}

The peculiar velocities of Class 1 LBVs are high on average, compared to the local fields. This is consistent with the high radial velocities found by \citet{Aghakhanloo2022}, and supports the premise of \citet{Smith2015}
that the majority of LBVs have been ejected from their birth clusters.

\begin{figure}
\plotone{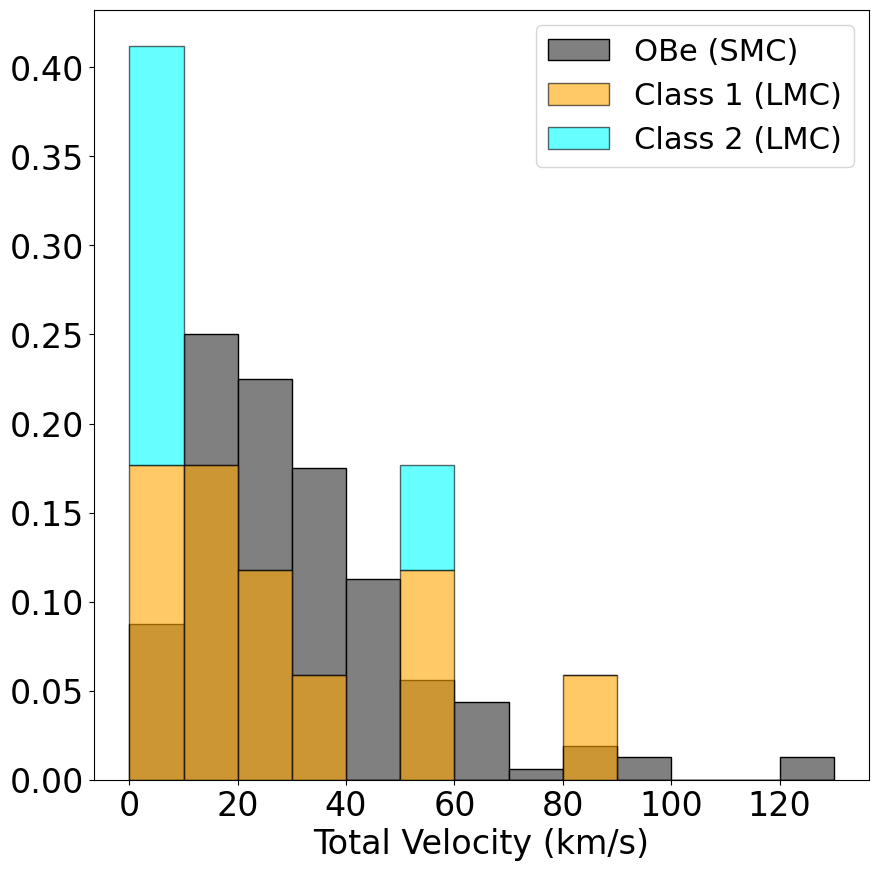}
\caption{
Normalized
comparison of LMC LBV stars with the SMC field OBe star sample.
The stacked contributions of Classes 1 and 2 are color-coded as shown.
}
\label{fig:LBVLBVc_OBehist}
\end{figure}

We can compare with the classical OBe stars, which are reasonably established to represent a population of ejected objects. These are believed to largely represent survivors of post-interaction binary mass transfer, accelerated by supernova kicks and/or orbital unbinding \citep[e.g.,][]{Dallas2022, Phillips2024}. Table~\ref{tab:LMCmedians} includes velocity statistics for the field SMC OBe stars from \citet{Phillips2024} that can be compared with the LMC LBV and sgB[e] populations. 
The peculiar velocities for these two samples are both measured using the same general algorithm that compares the target star's motion relative to the background field, as described earlier.
Figure~\ref{fig:LBVLBVc_OBehist} shows the normalized velocity distributions of the field SMC OBe stars and our LMC LBV sample.  

We caution that the OBe star sample consists of field objects from the RIOTS4 survey \citep{Lamb2016, Phillips2024}, which are all at least 28 pc from any OB or OBe star, while the LBV sample does not have this restriction and includes all available LBVs in the LMC as described above. However, \citet{Smith2015} find that LBVs 
have
median distances from the nearest O stars on the order of 40 -- 50 pc,
which may be compared to the above values for OBe field stars.

If LBVs are largely post-SN ejected objects, then we can expect their velocity distribution to be similar to that of the classical OBe stars.  The parameters for SN acceleration would likely differ between these two groups, so the velocity distributions are not expected to be identical, but there should be broad similarities, especially in shape.
Figure~\ref{fig:LBVLBVc_OBehist} shows 
the distributions for the LMC LBVs and
the SMC field OBe stars.  As seen above, the strong peak at the lowest velocities is due to the Class 2 objects, 
which seem largely unaccelerated.
When comparing only Class 1 LBVs with the OBe distribution, the two samples show substantial similarity, with the K-S test $p$-value at 0.59.
However, the Class 1 LBV distribution appears offset to lower velocities than the SMC OBe stars, and the characteristic velocities for Class 1 LBVs tend to be somewhat smaller than for the SMC OBe stars (Table~\ref{tab:LMCmedians}). This may be due to the omission of non-field SMC OBe stars; and/or, the effect could be real, since we expect LBVs to have much higher average masses than OBe stars.  Other physical or observational reasons could also be important.

Thus, we find that Class 1 LBVs have kinematics that are consistent with objects that have been ejected as runaways and walkaways from post-SN binaries, as suggested by \citet{Smith2015} and \citet{Aghakhanloo2017}. Our peculiar velocities are consistent with the results of
\citet{Aghakhanloo2022}, who find that over 33\% of LMC LBVs have radial velocities exceeding 25 km/s. In contrast, the Class 2 objects,
which are those lacking dusty nebulae,
appear to be unaccelerated, having a completely different velocity distribution from the Class 1 objects, as noted above (Figure~\ref{fig:LBV_SED}).  In agreement with the findings of \citet{Humphreys2016} and \citet{Aadland2018}, Class 2 objects therefore likely have a different origin, consistent with single-star evolution models.  Alternatively, they could be
pre-SN binaries, or
the product of binary mergers, which would experience very weak acceleration.  

\subsection{``Classical" and ``Less luminous" LBVs}\label{sec:classical}

\citet{Humphreys2016} suggest that ``classical" LBVs are not accelerated by binary SN explosions, thus it is also of interest to examine the stellar kinematics when using their LBV classification scheme for ``classical" and ``less luminous" objects. These authors identify S Dor, R127, and R143 as the only 
confirmed ``classical'' LBVs in the LMC, and they are known to be cluster members \citep{Smith2015}.  The ``classical" group is distinguished from ``less luminous" LBVs primarily by luminosity, and 
we caution that R143 has since been determined to have much lower luminosity than believed earlier, due to an apparent misidentification in the literature
\citep[Table~\ref{tab:vel_lmc};][]{Agliozzo2019}. 
We therefore drop this star as a ``classical" object.  We
instead include two stars that meet the \citet{Humphreys2016} luminosity criterion and are considered confirmed LBVs in Table~\ref{tab:vel_lmc}; since these were identified as only candidates by Humphreys et al., they did not include them in the ``classical" group.  These stars are R116 and HD 269582, which both belong to Class 2; while S Dor and R127 both belong to Class 1a. The remainder of the confirmed objects in Table~\ref{tab:vel_lmc} are the ``less luminous" LBVs.

\begin{figure}
\plotone{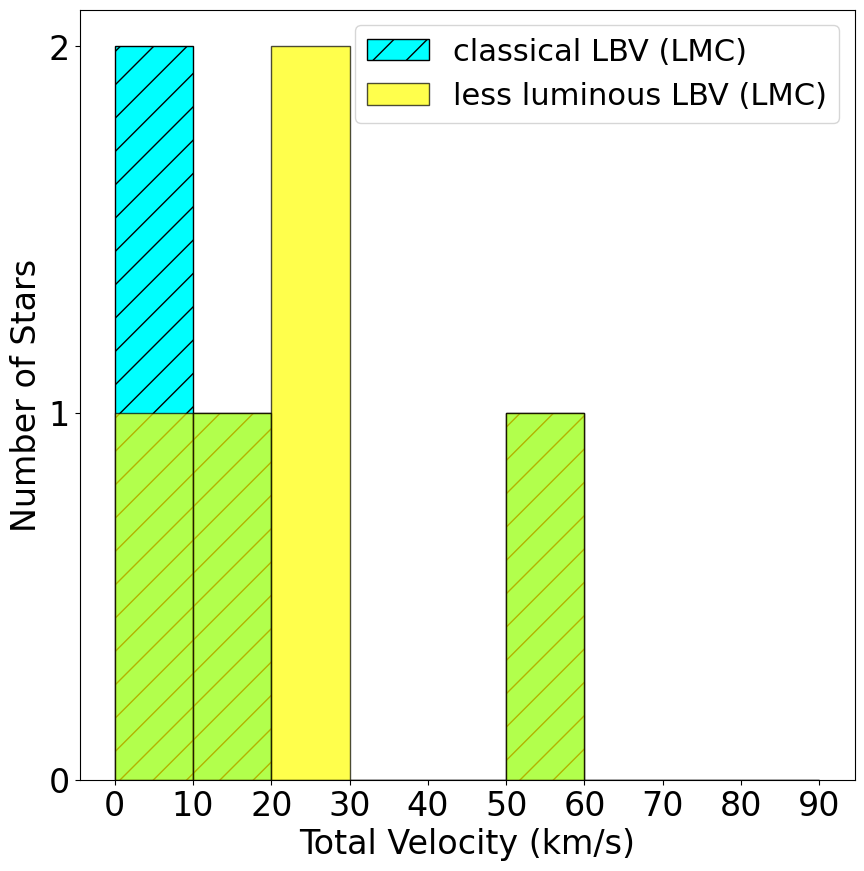}
\caption{
Velocity distributions for the ``classical" and ``less luminous" confirmed LBVs, as identified in Section~\ref{sec:classical}.  LBVc stars are not included. 
}
\label{fig:LBVclassical}
\end{figure}

Figure~\ref{fig:LBVclassical} shows the velocity distributions for these subsamples. We see that the ``less luminous" LBVs have higher velocities than the ``classical" objects.  Taking luminosity as a proxy for mass, this is consistent with the scenario where the most massive stars have the slowest velocities.  In the SN acceleration scenario, this effect is expected due to the slower acceleration expected for more massive stars.  
However, this does not preclude other scenarios to explain the ``classical" LBVs, including single-star evolution.

Figure~\ref{fig:scatterLBVLMC} indicates the ``classical'' objects with black dots on their symbols. We see that the Class 1 ``classical" LBVs have among the lowest transverse velocities for Class 1 objects.
As noted earlier, the definition of Class 2 objects, which have no dusty nebula and no dust emission, appears to be most closely related to the LBVs defined as ``classical" objects \citep{Humphreys2017b}.  And, we find that the 
peculiar velocities of the ``classical" LBVs are indeed slower, thus similar to the Class~2 objects. 
However, Figure~\ref{fig:scatterLBVLMC} shows that the velocity distribution for ``classical" LBVs is not as clearly limited to values $< 10\ \kms$, as seems to be the case for the Class 2 stars.
Thus the kinematics do not clarify whether there is a fundamental distinction between the ``classical" LBVs and the ``less luminous" objects.

\section{sgB[e] velocity distribution}

\subsection{Class 3 ``LBV candidate" stars}\label{sec:class3}

Turning now to sgB[e] stars, we first review the status of the Class 3 objects from \citet{Agliozzo2021}.
These objects have IR SEDs dominated by hot dust, and all of them
are also identified as sgB[e] stars in the literature; 
such dust is a defining feature of classical sgB[e] stars. We note that \citet{Agliozzo2021} based their sample on that of \citet{Richardson2018}, who focused only on identifying all possible candidate LBV stars without attempting to distinguish them from sgB[e] stars. The Class 3 group shows observational commonalities between LBV and sgB[e] stars. In particular, Table~\ref{tab:LMCluminosities} shows that their luminosities are comparable to Class 1 and 2 LBVs, whereas non-Class 3 sgB[e] stars have median luminosities that are an order of magnitude fainter.

\begin{figure}
\plotone{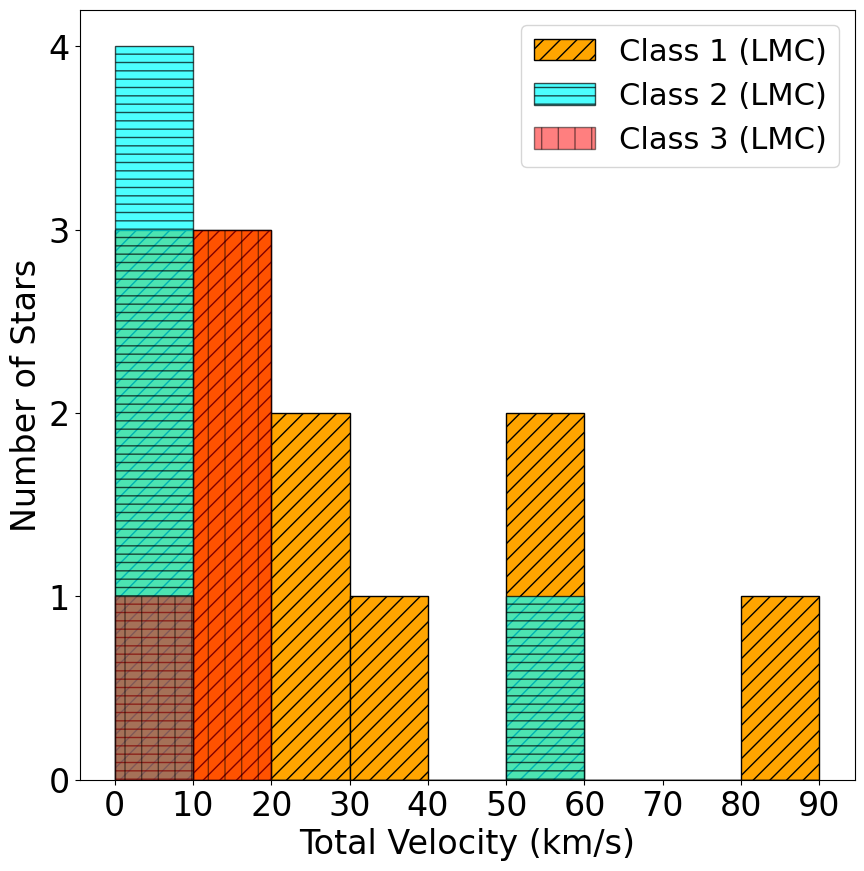}
\caption{Comparison of velocity distributions for Classes 1, 2, and 3 in the LMC. 
}
\label{fig:LBV123}
\end{figure}

Comparing the kinematics of Class 3 stars to the Class 1 and 2 LBV groups, there are some notable differences, as seen in Figure~\ref{fig:LBV123}. Class 3 stars all have relatively low velocities, $< 20$ km/s. Table~\ref{tab:LMCmedians} shows that the weighted median of Class 3 stars is significantly lower than that for the Class 1 objects. The Class 3 velocities are more similar to those of the Class 2 stars, which are also all slow, except for the single outlier.  However, the Class 2 objects are even slower, and consistent with having unaccelerated velocities, whereas the Class 3 kinematics are dominated by objects in the 10 -- 20 $\kms$ range, and thus consistent with origins as slow ejections, known as walkaway stars. 

\begin{figure}
\plotone{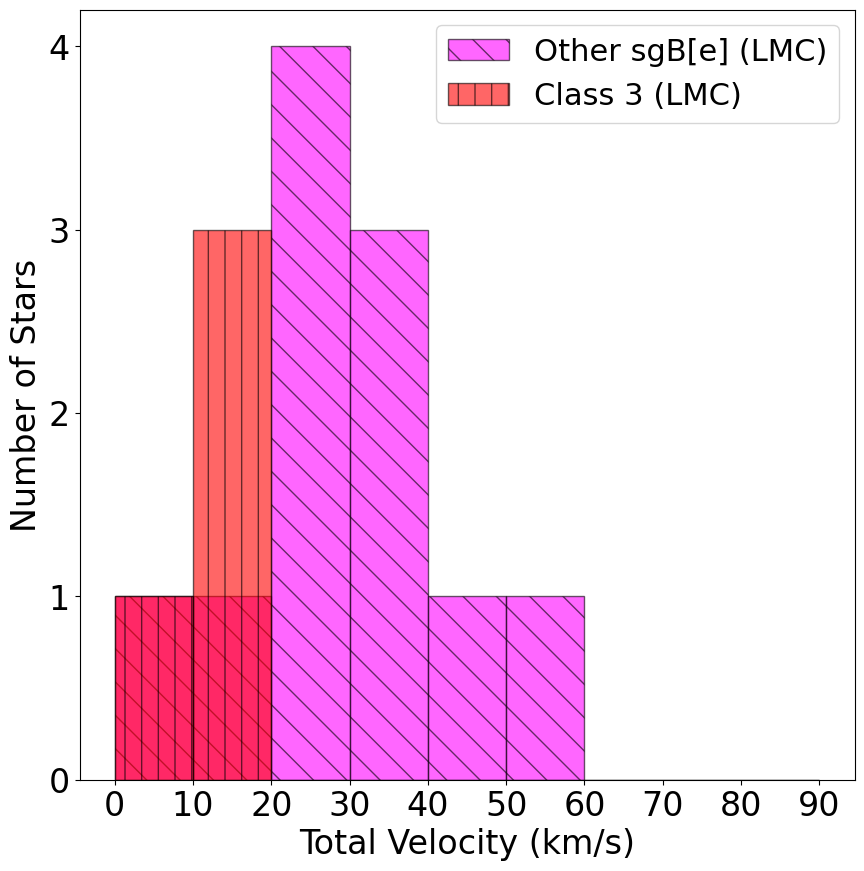}
\caption{Comparison of velocity distributions for Class 3 and the remaining sgB[e] stars in the LMC.}
\label{fig:SED3sgBe}
\end{figure}

\begin{figure}
\plotone{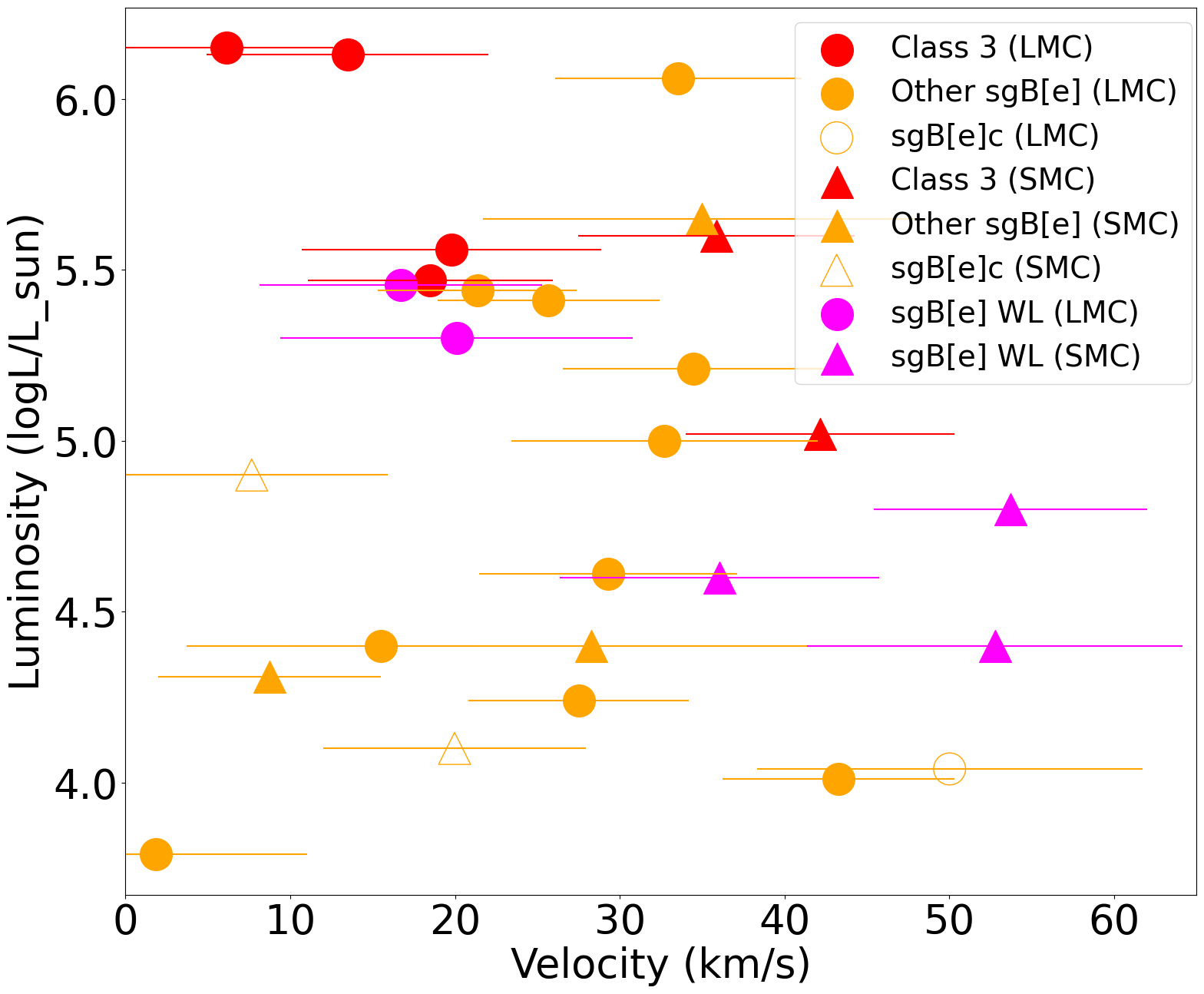}
\caption{ 
Luminosity vs velocity for 
Class 3 stars, non-Class 3 classical sgB[e] stars,
and sgB[e]-WL stars, color coded as shown. Circles and triangles indicate LMC and SMC objects, respectively.
}
\label{fig:scattersgBeLMC}
\end{figure}

Figure~\ref{fig:SED3sgBe} compares the Class 3 velocity distribution to that of other sgB[e] stars. The Class 3 stars are confined to slow velocities, and none are faster than $20\ \kms$.  In contrast,
the other sgB[e] stars have high velocities, peaking at 20--30 $\kms$, with a tail up to 60 $\kms$. The aggregate differences between these groups are also apparent in Table~\ref{tab:LMCmedians}. Figure~\ref{fig:scattersgBeLMC} shows that Class 3 objects are slower and more luminous than the non-Class 3 sgB[e] stars, with little overlap in their parameter space. We also see that the non-Class 3 sgB[e] stars have a wide range of luminosities and velocities, but their average velocities are higher than all LBV classes (Table~\ref{tab:LMCmedians}). 

Figures~\ref{fig:SED3sgBe} and \ref{fig:scattersgBeLMC} show that the luminosities and velocities of the LMC Class 3 objects form a continuous distribution with those of the remaining sgB[e] stars.
Since all classical sgB[e] stars also have strong dust emission,
the Class 3 stars appear to correspond to the most luminous and massive extreme of the of the sgB[e] population. This is consistent with their status as confirmed sgB[e] stars in the catalog of \citet{Kraus2019}.  We therefore  consider them to be distinct from LBVs.  We retain them all in our sgB[e] sample, and consider LBVs to include only Classes 1 and 2.

\subsection{sgB[e] stars vs LBVs}

\begin{figure}
\plotone{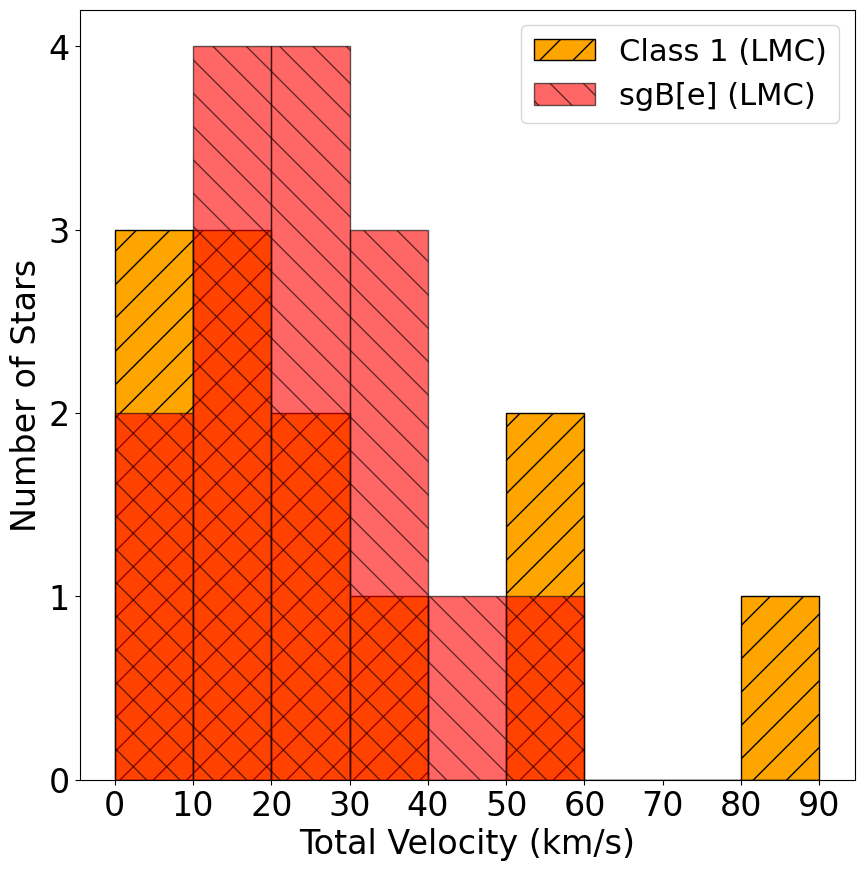}
\caption{
Velocity distributions 
of Class 1 LBVs and sgB[e] stars in the LMC.}
\label{fig:LBV1_sgBeLMC}
\end{figure}

Figure~\ref{fig:LBV1_sgBeLMC}
shows the velocity distribution for this combined LMC sgB[e] sample, showing that it is a fast population.
Since we found above that the Class~2 LBVs are largely unaccelerated,
we compare the the sgB[e] stars to only the
Class 1 LBVs in Figure~\ref{fig:LBV1_sgBeLMC}.  We see that the two distributions are qualitatively similar, which is consistent with the K-S test $p$-value of 0.72. The 
weighted median for Class 1 LBVs is 20 $\kms$, while the that for sgB[e] stars is 26 $\kms$. 

\begin{figure}
\plotone{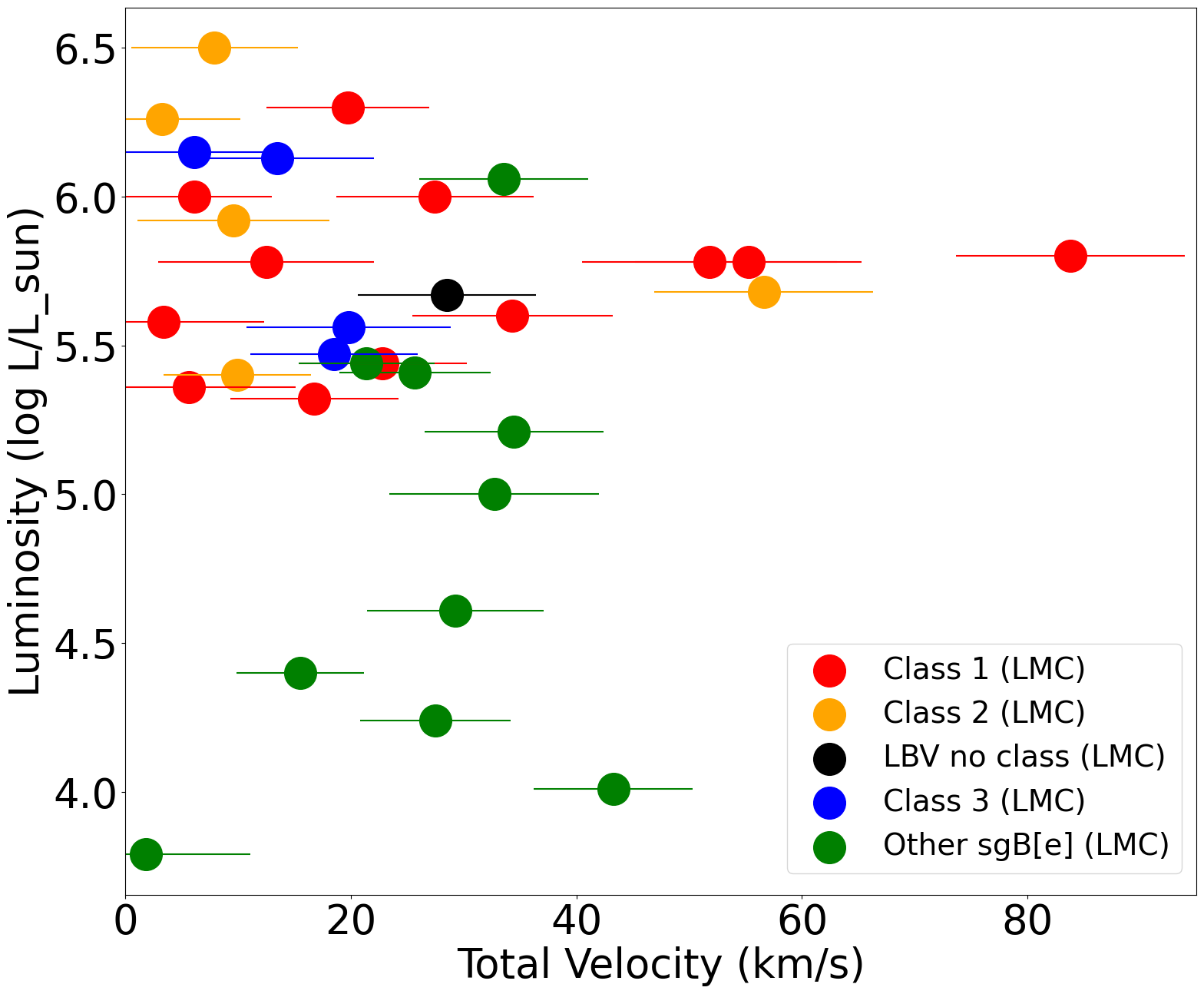}
\caption{
Luminosity vs velocity for LMC LBV Classes 1 and 2, and sgB[e] Class 3 and non-Class 3 objects, color-coded as shown.}
\label{fig:LMCscatter}
\end{figure}

These results suggest that Class 1 LBVs and sgB[e] stars may have similar kinematic origins. Class 1 LBVs are on average
much more luminous, and therefore more massive, than sgB[e] stars (Figure~\ref{fig:LMCscatter} and Table~\ref{tab:LMCluminosities}), which is qualitatively consistent with the difference in their mean velocities if both types of objects are accelerated by SN ejections. However, the substantial overlap in parameter space 
between the Class 1 LBVs and the combined sgB[e] classes
(Figure~\ref{fig:LMCscatter})
suggests that there are likely fundamental differences between these two populations as suggested by, e.g., \citet{Humphreys2017a}, rather than that sgB[e] stars are simply lower-mass analogs to LBV stars as suggested by other authors \citep[e.g.,][]{Smith2015}. 

We see that the Class 1 LBV population does include a few high-velocity outliers. It may be that these LBVs are accelerated by dynamical ejections.  Possibly they could result from two-step ejections combining both the dynamical and SN mechanisms. It is interesting that such high-velocity objects are not apparent among the sgB[e] stars, but additional data is needed to determine whether this difference is significant.  

\subsection{sgB[e] stars as possible binary products}\label{sec:sgBebinary}

\begin{figure}
\plotone{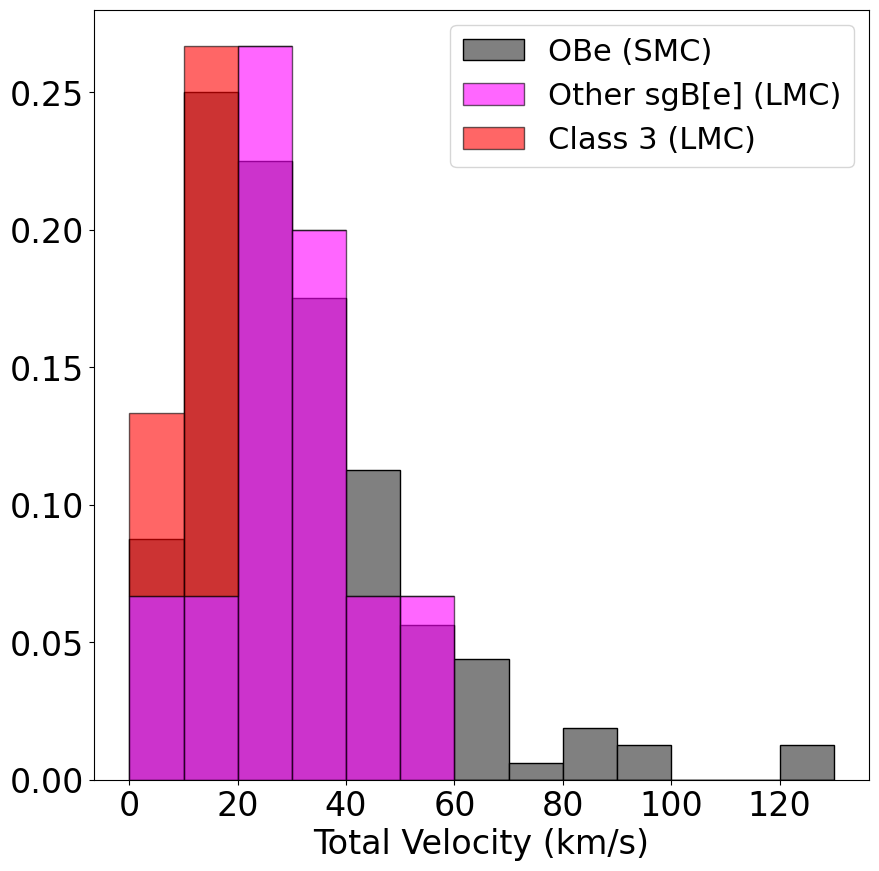}
\caption{Normalized velocity distributions of LMC sgB[e] stars and SMC field OBe stars, with Class 3  and non-Class 3 stars stacked and color-coded as shown.
}
\label{fig:sgBe_Obehist}
\end{figure}

Figure~\ref{fig:sgBe_Obehist} compares the normalized velocity distributions of the LMC sgB[e] stars and the SMC field OBe stars. The sgB[e] stars are known to be even more isolated than LBV stars, with essentially all of them being field objects \citep{Smith2015, Humphreys2017a, Kraus2019}. The distributions of sgB[e] and field OBe stars have strong similarities, including similar weighted medians  ($26\ \kms$ and $24\ \kms$, respectively; Table~\ref{tab:LMCmedians}).  This supports the scenario that sgB[e] stars are also accelerated by SN ejections, 
particularly since the OBe population excludes non-field stars.

{
There is significant evidence that sgB[e] stars are binary systems and that the B[e] phenomenon in these objects originates from a circumbinary disk \citep[see, e.g., review by][and references therein]{deWit2014}.
In a scenario where the sgB[e] stars are ejected by SNe in binary systems, any bound binary companions must be neutron stars or black holes.
Interestingly, detection of the inferred binary companions of sgB[e] stars is
often elusive, especially in the LMC and SMC.
In particular, only one of our sample sgB[e] stars, HD 38489 \citep{Bartlett2015}, is a known binary \citep{Kraus2019}.  If we also consider SMC sgB[e] stars,
LHA 115-S6 \citep[R4;][]{Zickgraf1996, Massey2014} and LHA 115-S18 \citep{Clark2013} are additional binaries, although LHA 115-S6 is also a candidate merger system \citep{Pasquali2000}.  The other two objects, HD 38489 and LHA 115-S18, are identified as binaries by their X-ray emission, attributed to colliding winds \citep{Clark2013}. These data are ambiguous enough that there may be a possibility that most sgB[e] stars in our sample might have neutron star or black hole companions.  }

Around 15\% of post-SN massive binaries remain bound in models by \citet{Renzo2019}, and their predicted characteristic velocities for these systems are around $20\ \kms$ for neutron stars and $10\ \kms$ for black holes. The latter value depends strongly on the treatment of fall-back from the explosion and could be higher.  We can compare these predictions with the observed sgB[e] velocity distribution in Figure~\ref{fig:LBV1_sgBeLMC}, keeping in mind that this shows 2-d projected velocities, which are on average smaller than the 3-d space velocities by a factor of 0.82.  The observed velocities are consistent with a possible scenario that most sgB[e] stars in the LMC and SMC are post-SN bound systems.

\section{SMC populations}

We now consider the SMC populations.  Table~\ref{tab:LMCmedians} shows 
characteristic velocity data for the different object classes. 
Given the small numbers of LBVs and sgB[e] stars in this galaxy, interpretations of their data are inconclusive; however, observed variations could be suggestive of effects due to the SMC's lower metallicity.

\subsection{LBV stars}

There are two LBV stars identified in the SMC.  According to \citet{Agliozzo2021}, HD 6884 belongs to Class 1b, and HD 5980 to Class 2. 
HD 6884 has a low velocity (11.6$\pm{6.5}$ $\kms$), which is not necessarily unusual for Class 1 LBVs, as shown in Figures~\ref{fig:scatterLBVLMC} and \ref{fig:LBV123}. On the other hand, HD 5980 is a complicated and unusual multiple system, which includes a Wolf-Rayet star, and its photometric variability is attributed to both hydrostatic instability and binary interactions \citep{Koenigsberger2014, Hillier2019}. Therefore, this object may not be a true LBV \citep{Humphreys2016}. We find that it is traveling at high speed (55.3$\pm{14.1}$ $\kms$), which is unusual for Class~2 objects in the LMC, although as seen in  Figures~\ref{fig:scatterLBVLMC} and \ref{fig:LBV123}, one other Class 2 LBV does have a similar velocity. Without more stars in the SMC sample, it is hard to evaluate whether there are significant differences from the LMC trends. 

\subsection{sgB[e] stars}

There are 7 sgB[e] stars in the SMC sample, including the 2 sgB[e]c stars.
The sample also includes 2 Class 3 stars identified by \citet{Agliozzo2021}, LHA 115-S 6 and LHA 115-S 18.  
Similar to the LMC Class 3 objects, both are established in the sgB[e] census of \citet{Kraus2019}.  We see that these SMC Class 3 stars are faster than the other sgB[e] stars (Tables~\ref{tab:vel_smc} and \ref{tab:LMCmedians}), a trend opposite to what is seen in the LMC (Figure~\ref{fig:scattersgBeLMC}).
However, as discussed above in Section~\ref{sec:sgBebinary}, both of these stars are likely non-compact binaries, which therefore can only be accelerated by the dynamical process.  Dynamical ejections, including runaway binaries, typically attain higher velocities than SN ejections, as has been observed in the SMC \citep{Phillips2024, Oey2018}, including at values shown by the Class 3 objects.
These Class 3 stars still have higher luminosities than the other sgB[e] stars, consistent with what we found above (Section~\ref{sec:class3}) for the LMC sgB[e] stars, although the two SMC Class 3 stars have somewhat lower luminosities than their LMC counterparts. The high velocities of the SMC objects suggest that dynamical processes may play a role in their kinematics.

\begin{figure}
\plotone{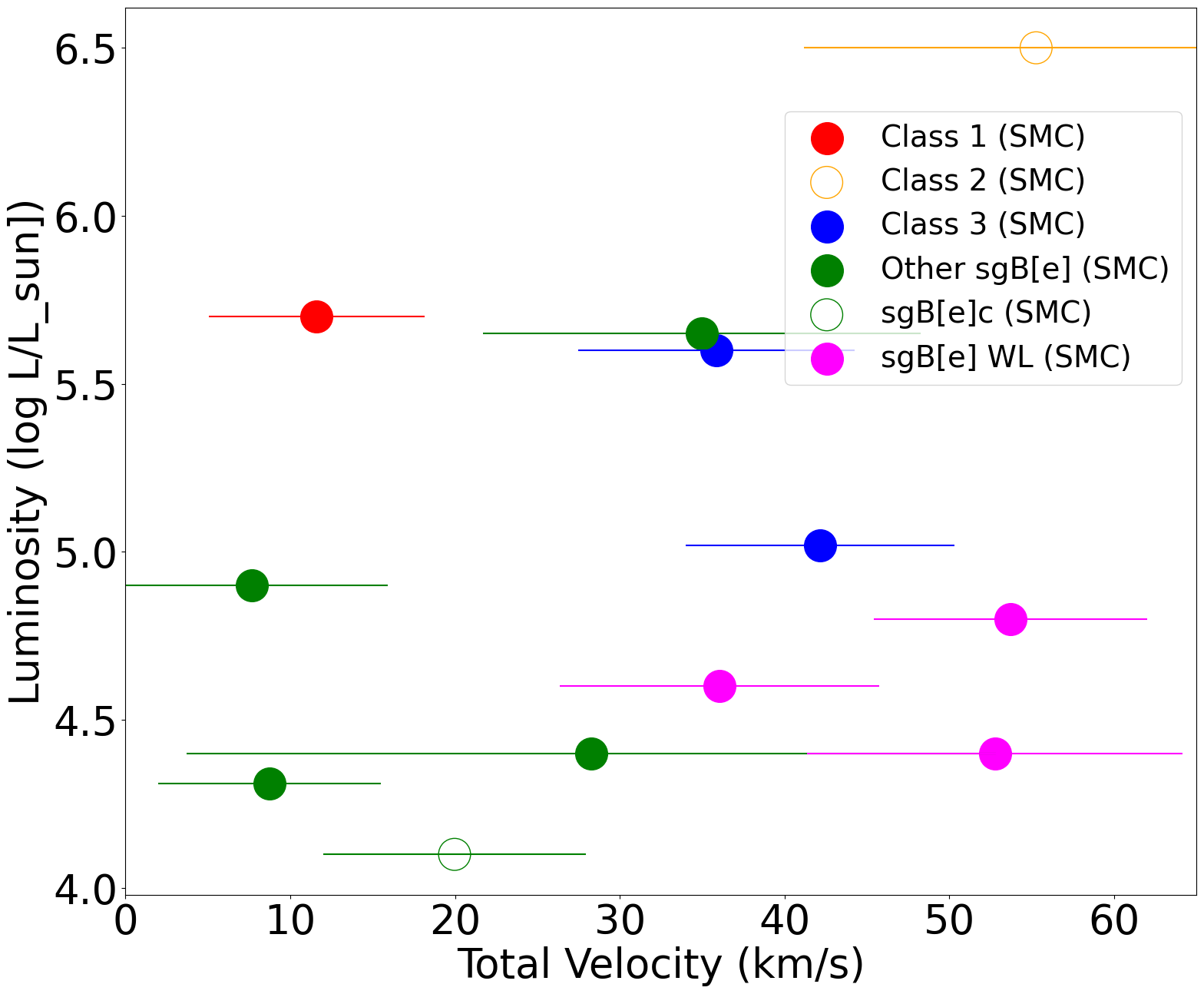}
\caption{
Luminosity vs velocity for all stars in the SMC.  
LBVc stars are shown with open circles;
the Class 2 object is HD 5980, whose LBV status is especially questionable.  }
\label{fig:scatterSMC}
\end{figure}

\begin{figure}
\plotone{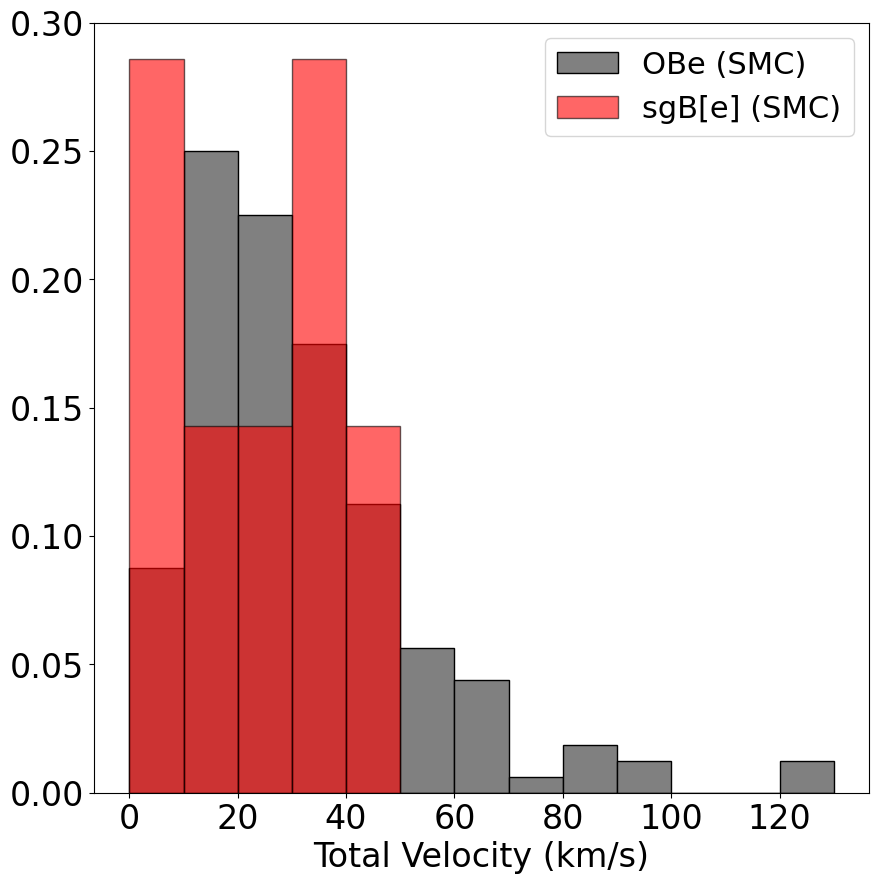}
\caption{
Normalized velocity distributions for
the sgB[e] and field OBe stars in the SMC.}
\label{fig:OBesgBeSMC}
\end{figure}

Figure~\ref{fig:OBesgBeSMC} compares the velocities of the 7 sgB[e] stars of the SMC with the field OBe stars. This now has the advantage of comparing populations within the same galaxy. The velocity distribution of the sgB[e] stars appears to be flatter than for the OBe stars, but the data are noisy and the KS-test $p$-value is 0.82, showing that the difference is not statistically significant. In any case, we see that the data are certainly consistent with most SMC sgB[e] stars also being 
accelerated and ejected from their parent clusters.
If the contrasting velocity distributions are confirmed, the SMC sgB[e] objects may experience stronger SN kicks or a greater contribution from dynamical ejection processes.

\section{Weak-Lined sgB[e] Stars}

As mentioned earlier, there is a distinct subset of dust-poor sgB[e] stars identified by \citet{Graus2012}, which we refer to as weak-lined sgB[e] (hereafter sgB[e]-WL) stars, due to their much weaker forbidden emission lines compared to other sgB[e] stars. \citet{Graus2012} find that while
sgB[e]-WL stars have no mid-IR dust emission, they do show free-free emission, which is also seen in the Class 2 LBVs and OBe stars. This indicates that the circumstellar emission is dominated by ionized gas.
These objects are different enough from standard sgB[e] stars that \citet{Kraus2019} considers them to be ``erroneous classifications", although historically the B[e] classification is based only on the presence of optical forbidden emission features, as discussed by \citet{Graus2012} \citep[see, e.g.,][]{Conti1976, Lamers1998}.

\citet{Graus2012} identify SMC stars LHA 115-S 29, LHA 15-S 46, and LHA 115-S 62 as members of this group, along with the LMC star VFTS 698. 
We now also include 
the LMC star {[L72]} LH 85-10 
as a member of this class, based on its IR SED, which is dominated by free-free emission instead of dust \citep{Bonanos2009}.
Due to the very small sample size of sgB[e]-WL stars, it is again difficult to make any conclusive interpretations of their kinematics.  

\begin{figure}
\plotone{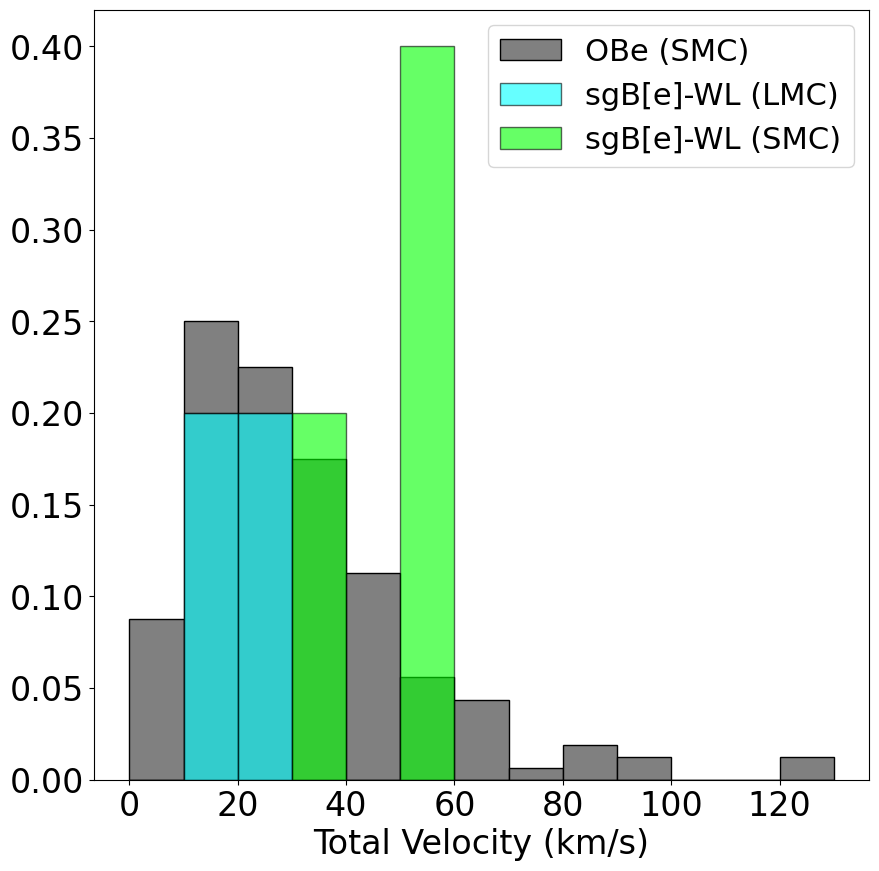}
\caption{
Velocity distributions for the sgB[e]-WL stars and the SMC field OBe stars.
Blue and green correspond to LMC and SMC objects, respectively.
} 
\label{fig:OBesgBeWL}
\end{figure}

Figure~\ref{fig:OBesgBeWL} shows the velocity distributions of the sgB[e]-WL stars for both galaxies compared to the SMC field OBe stars.  
The SMC sgB[e]-WL stars have very high velocities, exceeding those of any other group (Table~\ref{tab:LMCmedians}), whereas the two LMC sgB[e]-WL stars are slower. 
However, all of these objects again show accelerated proper motion velocities. 

In the SMC, their kinematics reinforce previous conclusions that the sgB[e]-WL stars have little in common with the other groups, including other sgB[e] stars, and should be treated as a distinct phenomenon. 
We see that their velocities are by far the fastest of any group (Figure~\ref{fig:scattersgBeLMC} and Table~\ref{tab:LMCmedians}).
Their luminosities imply somewhat lower masses, which promote faster SN ejection velocities, but the observed velocities, having a weighted median of 53 $\kms$. imply very strong kick velocities and/or a significant contribution from dynamical processes for all 3 objects.  

The LMC objects do not share these distinctive kinematics, and
the two sgB[e]-WL stars have velocities similar to those of normal sgB[e] stars and Class~1 LBVs (Figure~\ref{fig:scattersgBeLMC} and Table~\ref{tab:LMCmedians}).  If the LMC objects are indeed true analogs of their SMC counterparts, their slower velocities may in part be linked to their higher luminosities, and presumably, higher masses.  The lower SMC metallicity may also play an important role, since under these conditions, massive interacting binaries experience less orbital widening and are therefore tighter, leading to faster ejections \citep{Renzo2019}. 
However, we again caution that the small number statistics make any interpretation difficult.

\section{Conclusions}

The origin of LBV and sgB[e] stars has been debated for decades, with traditional single-star models depicting LBVs as a transitional class that occurs after early O stars lose their outer layers to evolve into WR stars.  The sgB[e] stars have often been considered to be closely related to LBVs. 
On the other hand, origins linked to binary interactions have also long been considered.  Since binary interactions usually lead to post-SN acceleration, the stellar kinematics provide crucial insight on these models.  In this paper, we carry out this test 
by measuring the peculiar proper motions for the LBVs and sgB[e] stars in the LMC and SMC.  

We focus primarily on the LMC, where
we find that the velocities of LBVs and LBVc stars are statistically indistinguishable, suggesting that they are indeed a single general group. We consider the IR classes of \citet{Agliozzo2021}, where Class 1 objects have dusty nebulae,
pointing to likely previous mass ejection; these
objects have velocity distributions that are comparable to those another group of massive stars known to be kicked from binaries, the SMC field OBe stars.  This suggests that Class 1
LBVs are also kicked by supernovae in binary systems in both the LMC and SMC, consistent with the widely dispersed spatial distribution reported by \citet{Smith2015}.

On the other hand, the Class 2 LBV stars,
which lack dusty nebulae,
generally show no evidence of acceleration.  They are also more luminous than the Class 1 stars, and therefore more massive and harder to accelerate via the supernova mechanism. However, the kinematics suggest that it is more likely that they are a fundamentally different class of objects with a different origin 
or evolutionary state, perhaps being pre-SN binaries or merger products.
While the criteria for identifying ``classical" LBVs differ significantly from those that identify Class 2 LBVs, these classifications may be physically related;
the kinematics of Class 2 objects are consistent with the findings by \citet{Humphreys2016} that ``classical" LBVs remain in clusters, which is also consistent with their being products of single-star evolution.

The overall velocity distribution of sgB[e] stars is also similar to that of the SMC field OBe stars, with higher average velocities than the LBV Class 1 stars.  This may be attributed to the lower masses of the sgB[e] stars.  Thus, the kinematics of sgB[e] stars are also consistent with binary supernova origins.  The sgB[e] sample includes the Class 3 stars of \citet{Agliozzo2021}, which they consider to be LBVc stars, but all are identified as sgB[e] stars in the census of \citet{Kraus2019}. 
The Class 3 stars are defined by their dusty spectra, which is also a defining property of classical sgB[e] stars. We find that the Class 3 stars and the remaining LMC sgB[e] stars form a continuous distribution in luminosity and velocity, fully consistent with the Class 3 objects simply corresponding to the most luminous and massive extreme of the sgB[e] population, and we therefore treat them as such.  

Although sgB[e] stars extend to much lower luminosities and higher velocities than Class 1 LBVs, there is substantial overlap between them in this parameter space, implying that the two types of objects are distinct, and that sgB[e] stars are not simply lower luminosity LBVs.  This is also supported by the clear difference in dust properties \citep[e.g.,][]{Kraus2019}.

We also obtain the peculiar velocities of SMC LBVs and sgB[e] stars.  Not much can be inferred for the LBV stars, since there are only two objects, one of which is unusual.  The SMC sgB[e] stars, even the most luminous ones, tend to have substantially higher velocities than the LMC counterparts.  This may be consistent with the expectation that lower-metallicity binaries are tighter, leading to faster ejection velocities \citep{Renzo2019}.  There may also be greater contribution from dynamical ejections, especially for the most massive objects.

We also examine the weak-lined sgB[e] stars \citep{Graus2012}, which are distinct from classical sgB[e] stars in that they show only free-free emission and no dust in the IR. All of these objects also show accelerated velocities. The SMC objects are again much faster than their LMC counterparts, and they have the fastest velocities of any group studied here.  This could be due to the same metallicity effect mentioned above.  Another possible contributing factor might be that the SMC objects may be overluminous, lower-mass objects.
 
Thus, the kinematics of the Class 1 LBVs, sgB[e] stars, and sgB[e]-WL stars all show accelerated velocities, implying that these objects generally are ejected from their parent clusters.  The velocity distributions of the Class 1 LBVs and sgB[e] stars are similar in form to that of the SMC field OBe stars, which are expected to be products of binary mass transfer.  Thus it is likely that the LBVs
with dusty nebulae
and sgB[e] stars have similar origins.  However, a significant contribution from dynamical ejections cannot be ruled out, since the timescale for this process is on the order of the stellar lifetimes \citep[e.g.,][]{Brinkmann2017}. 
On the other hand, the Class 2 LBVs, which are luminous objects 
with no dusty nebulae,
generally show no evidence of acceleration and may 
form according to single-star evolution models, or they could be pre-SN binaries or binary merger products \citep[e.g.,][]{Aghakhanloo2017}.

\begin{acknowledgments}
We gratefully thank all who discussed this work with us, 
including Mojgan Aghakhanloo, Johnny Dorigo Jones, Grant Phillips, Mathieu Renzo, and Irene Vargas-Salazar.
We also thank the referee for comments and suggestions that substantially improved this paper.
\end{acknowledgments}


\vspace{5mm}
\facilities{Gaia, SIMBAD, Vizier}


\software{Astropy \citep{Astropy},  
Scipy \citep{Scipy}
          }

\bibliography{LBVcitation.bib}{}

\begin{thebibliography}{}
\expandafter\ifx\csname natexlab\endcsname\relax\def\natexlab#1{#1}\fi
\providecommand{\url}[1]{\href{#1}{#1}}
\providecommand{\dodoi}[1]{doi:~\href{http://doi.org/#1}{\nolinkurl{#1}}}
\providecommand{\doeprint}[1]{\href{http://ascl.net/#1}{\nolinkurl{http://ascl.net/#1}}}
\providecommand{\doarXiv}[1]{\href{https://arxiv.org/abs/#1}{\nolinkurl{https://arxiv.org/abs/#1}}}

\bibitem[{{Aadland} {et~al.}(2018){Aadland}, {Massey}, {Neugent}, \&
  {Drout}}]{Aadland2018}
{Aadland}, E., {Massey}, P., {Neugent}, K.~F., \& {Drout}, M.~R. 2018, \aj,
  156, 294, \dodoi{10.3847/1538-3881/aaeb96}

\bibitem[{{Aghakhanloo} {et~al.}(2017){Aghakhanloo}, {Murphy}, {Smith}, \&
  {Hlo{\v{z}}ek}}]{Aghakhanloo2017}
{Aghakhanloo}, M., {Murphy}, J.~W., {Smith}, N., \& {Hlo{\v{z}}ek}, R. 2017,
  \mnras, 472, 591, \dodoi{10.1093/mnras/stx2050}

\bibitem[{{Aghakhanloo} {et~al.}(2022){Aghakhanloo}, {Smith}, {Andrews},
  {Olsen}, {Besla}, \& {Choi}}]{Aghakhanloo2022}
{Aghakhanloo}, M., {Smith}, N., {Andrews}, J., {et~al.} 2022, \mnras, 516,
  2142, \dodoi{10.1093/mnras/stac2265}

\bibitem[{{Agliozzo} {et~al.}(2019){Agliozzo}, {Mehner}, {Phillips}, {Leto},
  {Groh}, {Noriega-Crespo}, {Buemi}, {Cavallaro}, {Cerrigone}, {Ingallinera},
  {Paladini}, {Pignata}, {Trigilio}, \& {Umana}}]{Agliozzo2019}
{Agliozzo}, C., {Mehner}, A., {Phillips}, N.~M., {et~al.} 2019, \aap, 626,
  A126, \dodoi{10.1051/0004-6361/201935239}

\bibitem[{{Agliozzo} {et~al.}(2021){Agliozzo}, {Phillips}, {Mehner}, {Baade},
  {Scicluna}, {Kemper}, {Asmus}, {de Wit}, \& {Pignata}}]{Agliozzo2021}
{Agliozzo}, C., {Phillips}, N., {Mehner}, A., {et~al.} 2021, \aap, 655, A98,
  \dodoi{10.1051/0004-6361/202141279}

\bibitem[{{Astropy Collaboration} {et~al.}(2013){Astropy Collaboration},
  {Robitaille}, {Tollerud}, {Greenfield}, {Droettboom}, {Bray}, {Aldcroft},
  {Davis}, {Ginsburg}, {Price-Whelan}, {Kerzendorf}, {Conley}, {Crighton},
  {Barbary}, {Muna}, {Ferguson}, {Grollier}, {Parikh}, {Nair}, {Unther},
  {Deil}, {Woillez}, {Conseil}, {Kramer}, {Turner}, {Singer}, {Fox}, {Weaver},
  {Zabalza}, {Edwards}, {Azalee Bostroem}, {Burke}, {Casey}, {Crawford},
  {Dencheva}, {Ely}, {Jenness}, {Labrie}, {Lim}, {Pierfederici}, {Pontzen},
  {Ptak}, {Refsdal}, {Servillat}, \& {Streicher}}]{Astropy}
{Astropy Collaboration}, {Robitaille}, T.~P., {Tollerud}, E.~J., {et~al.} 2013,
  \aap, 558, A33, \dodoi{10.1051/0004-6361/201322068}

\bibitem[{{Bartlett} \& {Clark}(2015)}]{Bartlett2015}
{Bartlett}, E.~S., \& {Clark}, J.~S. 2015, in SALT Science Conference 2015
  (SSC2015), ed. D.~{Buckley} \& A.~{Schroeder}, 55,
  \dodoi{10.22323/1.250.0055}

\bibitem[{{Bonanos} {et~al.}(2009){Bonanos}, {Massa}, {Sewilo}, {Lennon},
  {Panagia}, {Smith}, {Meixner}, {Babler}, {Bracker}, {Meade}, {Gordon},
  {Hora}, {Indebetouw}, \& {Whitney}}]{Bonanos2009}
{Bonanos}, A.~Z., {Massa}, D.~L., {Sewilo}, M., {et~al.} 2009, \aj, 138, 1003,
  \dodoi{10.1088/0004-6256/138/4/1003}

\bibitem[{{Bouret} {et~al.}(2005){Bouret}, {Lanz}, \& {Hillier}}]{Bouret2005}
{Bouret}, J.~C., {Lanz}, T., \& {Hillier}, D.~J. 2005, \aap, 438, 301,
  \dodoi{10.1051/0004-6361:20042531}

\bibitem[{{Brinkmann} {et~al.}(2017){Brinkmann}, {Banerjee}, {Motwani}, \&
  {Kroupa}}]{Brinkmann2017}
{Brinkmann}, N., {Banerjee}, S., {Motwani}, B., \& {Kroupa}, P. 2017, \aap,
  600, A49, \dodoi{10.1051/0004-6361/201629312}

\bibitem[{{Clark} {et~al.}(2013){Clark}, {Bartlett}, {Coe}, {Dorda}, {Haberl},
  {Lamb}, {Negueruela}, \& {Udalski}}]{Clark2013}
{Clark}, J.~S., {Bartlett}, E.~S., {Coe}, M.~J., {et~al.} 2013, \aap, 560, A10,
  \dodoi{10.1051/0004-6361/201321216}

\bibitem[{{Conti}(1975)}]{Conti1975}
{Conti}, P.~S. 1975, Memoires of the Societe Royale des Sciences de Liege, 9,
  193

\bibitem[{{Conti}(1976)}]{Conti1976}
{Conti}, P.~S. 1976, in Be and Shell Stars, ed. A.~{Slettebak}, IAU Symposium
  70, 447

\bibitem[{{Conti}(1979)}]{Conti1979}
{Conti}, P.~S. 1979, in Mass Loss and Evolution of O-Type Stars, ed. P.~S.
  {Conti} \& C.~W.~H. {De Loore}, Vol.~83, 431--443

\bibitem[{{Dallas} {et~al.}(2022){Dallas}, {Oey}, \& {Castro}}]{Dallas2022}
{Dallas}, M.~M., {Oey}, M.~S., \& {Castro}, N. 2022, \apj, 936, 112,
  \dodoi{10.3847/1538-4357/ac8988}

\bibitem[{{de Koter} {et~al.}(1996){de Koter}, {Lamers}, \&
  {Schmutz}}]{Koter1996}
{de Koter}, A., {Lamers}, H.~J.~G.~L.~M., \& {Schmutz}, W. 1996, \aap, 306, 501

\bibitem[{{de Wit} {et~al.}(2014){de Wit}, {Oudmaijer}, \& {Vink}}]{deWit2014}
{de Wit}, W.~J., {Oudmaijer}, R.~D., \& {Vink}, J.~S. 2014, Advances in
  Astronomy, 2014, 270848, \dodoi{10.1155/2014/270848}

\bibitem[{{Dorigo Jones} {et~al.}(2020){Dorigo Jones}, {Oey}, {Paggeot},
  {Castro}, \& {Moe}}]{DorigoJones2020}
{Dorigo Jones}, J., {Oey}, M.~S., {Paggeot}, K., {Castro}, N., \& {Moe}, M.
  2020, \apj, 903, 43, \dodoi{10.3847/1538-4357/abbc6b}

\bibitem[{{Dunstall} {et~al.}(2012){Dunstall}, {Fraser}, {Clark}, {Crowther},
  {Dufton}, {Evans}, {Lennon}, {Soszy{\'n}ski}, {Taylor}, \&
  {Vink}}]{Dunstall2012}
{Dunstall}, P.~R., {Fraser}, M., {Clark}, J.~S., {et~al.} 2012, \aap, 542, A50,
  \dodoi{10.1051/0004-6361/201218872}

\bibitem[{{Gaia Collaboration} {et~al.}(2022){Gaia Collaboration},
  {Bailer-Jones}, {Teyssier}, {Delchambre}, {Ducourant}, {Garabato},
  {Hatzidimitriou}, {Klioner}, {Rimoldini}, {Bellas-Velidis}, {Carballo},
  {Carnerero}, {Diener}, {Fouesneau}, {Galluccio}, {Gavras}, {Krone-Martins},
  {Raiteri}, {Teixeira}, {Brown}, {Vallenari}, {Prusti}, {de Bruijne},
  {Arenou}, {Babusiaux}, {Biermann}, {Creevey}, {Evans}, {Eyer}, {Guerra},
  {Hutton}, {Jordi}, {Lammers}, {Lindegren}, {Luri}, {Mignard}, {Panem},
  {Pourbaix}, {Randich}, {Sartoretti}, {Soubiran}, {Tanga}, {Walton},
  {Bastian}, {Drimmel}, {Jansen}, {Katz}, {Lattanzi}, {van Leeuwen}, {Bakker},
  {Cacciari}, {Casta{\~n}eda}, {De Angeli}, {Fabricius}, {Fr{\'e}mat},
  {Guerrier}, {Heiter}, {Masana}, {Messineo}, {Mowlavi}, {Nicolas},
  {Nienartowicz}, {Pailler}, {Panuzzo}, {Riclet}, {Roux}, {Seabroke}, {Sordo},
  {Th{\'e}venin}, {Gracia-Abril}, {Portell}, {Altmann}, {Andrae}, {Audard},
  {Benson}, {Berthier}, {Blomme}, {Burgess}, {Busonero}, {Busso},
  {C{\'a}novas}, {Carry}, {Cellino}, {Cheek}, {Clementini}, {Damerdji},
  {Davidson}, {de Teodoro}, {Nu{\~n}ez Campos}, {Dell'Oro}, {Esquej},
  {Fern{\'a}ndez-Hern{\'a}ndez}, {Fraile}, {Garc{\'\i}a-Lario}, {Gosset},
  {Haigron}, {Halbwachs}, {Hambly}, {Harrison}, {Hern{\'a}ndez}, {Hestroffer},
  {Hodgkin}, {Holl}, {Jan{\ss}en}, {Jevardat de Fombelle}, {Jordan},
  {Lanzafame}, {L{\"o}ffler}, {Marchal}, {Marrese}, {Moitinho}, {Muinonen},
  {Osborne}, {Pancino}, {Pauwels}, {Recio-Blanco}, {Reyl{\'e}}, {Riello},
  {Roegiers}, {Rybizki}, {Sarro}, {Siopis}, {Smith}, {Sozzetti}, {Utrilla},
  {van Leeuwen}, {Abbas}, {{\'A}brah{\'a}m}, {Abreu Aramburu}, {Aerts},
  {Aguado}, {Ajaj}, {Aldea-Montero}, {Altavilla}, {{\'A}lvarez}, {Alves},
  {Anderson}, {Anglada Varela}, {Antoja}, {Baines}, {Baker},
  {Balaguer-N{\'u}{\~n}ez}, {Balbinot}, {Balog}, {Barache}, {Barbato},
  {Barros}, {Barstow}, {Bartolom{\'e}}, {Bassilana}, {Bauchet}, {Becciani},
  {Bellazzini}, {Berihuete}, {Bernet}, {Bertone}, {Bianchi}, {Binnenfeld},
  {Blanco-Cuaresma}, {Boch}, {Bombrun}, {Bossini}, {Bouquillon}, {Bragaglia},
  {Bramante}, {Breedt}, {Bressan}, {Brouillet}, {Brugaletta}, {Bucciarelli},
  {Burlacu}, {Butkevich}, {Buzzi}, {Caffau}, {Cancelliere}, {Cantat-Gaudin},
  {Carlucci}, {Carrasco}, {Casamiquela}, {Castellani}, {Castro-Ginard},
  {Chaoul}, {Charlot}, {Chemin}, {Chiaramida}, {Chiavassa}, {Chornay},
  {Comoretto}, {Contursi}, {Cooper}, {Cornez}, {Cowell}, {Crifo}, {Cropper},
  {Crosta}, {Crowley}, {Dafonte}, {Dapergolas}, {David}, {de Laverny}, {De
  Luise}, {De March}, {De Ridder}, {de Souza}, {de Torres}, {del Peloso}, {del
  Pozo}, {Delbo}, {Delgado}, {Delisle}, {Demouchy}, {Dharmawardena}, {Diakite},
  {Distefano}, {Dolding}, {Enke}, {Fabre}, {Fabrizio}, {Faigler}, {Fedorets},
  {Fernique}, {Figueras}, {Fournier}, {Fouron}, {Fragkoudi}, {Gai},
  {Garcia-Gutierrez}, {Garcia-Reinaldos}, {Garc{\'\i}a-Torres}, {Garofalo},
  {Gavel}, {Gerlach}, {Geyer}, {Giacobbe}, {Gilmore}, {Girona}, {Giuffrida},
  {Gomel}, {Gomez}, {Gonz{\'a}lez-N{\'u}{\~n}ez},
  {Gonz{\'a}lez-Santamar{\'\i}a}, {Gonz{\'a}lez-Vidal}, {Granvik}, {Guillout},
  {Guiraud}, {Guti{\'e}rrez-S{\'a}nchez}, {Guy}, {Hauser}, {Haywood}, {Helmer},
  {Helmi}, {Sarmiento}, {Hidalgo}, {H{\l}adczuk}, {Hobbs}, {Holland}, {Huckle},
  {Jardine}, {Jasniewicz}, {Jean-Antoine Piccolo}, {Jim{\'e}nez-Arranz},
  {Juaristi Campillo}, {Julbe}, {Karbevska}, {Kervella}, {Khanna}, {Kontizas},
  {Kordopatis}, {Korn}, {K{\'o}sp{\'a}l}, {Kostrzewa-Rutkowska},
  {Kruszy{\'n}ska}, {Kun}, {Laizeau}, {Lambert}, {Lanza}, {Lasne}, {Le
  Campion}, {Lebreton}, {Lebzelter}, {Leccia}, {Leclerc}, {Lecoeur-Taibi},
  {Liao}, {Licata}, {Lindstr{\o}m}, {Lister}, {Livanou}, {Lobel}, {Lorca},
  {Loup}, {Madrero Pardo}, {Magdaleno Romeo}, {Managau}, {Mann}, {Manteiga},
  {Marchant}, {Marconi}, {Marcos}, {Marcos Santos}, {Mar{\'\i}n Pina},
  {Marinoni}, {Marocco}, {Marshall}, {Polo}, {Mart{\'\i}n-Fleitas}, {Marton},
  {Mary}, {Masip}, {Massari}, {Mastrobuono-Battisti}, {Mazeh}, {McMillan},
  {Messina}, {Michalik}, {Millar}, {Mints}, {Molina}, {Molinaro}, {Moln{\'a}r},
  {Monari}, {Mongui{\'o}}, {Montegriffo}, {Montero}, {Mor}, {Mora},
  {Morbidelli}, {Morel}, {Morris}, {Muraveva}, {Murphy}, {Musella}, {Nagy},
  {Noval}, {Oca{\~n}a}, {Ogden}, {Ordenovic}, {Osinde}, {Pagani}, {Pagano},
  {Palaversa}, {Palicio}, {Pallas-Quintela}, {Panahi}, {Payne-Wardenaar},
  {Pe{\~n}alosa Esteller}, {Penttil{\"a}}, {Pichon}, {Piersimoni}, {Pineau},
  {Plachy}, {Plum}, {Poggio}, {Pr{\v{s}}a}, {Pulone}, {Racero}, {Ragaini},
  {Rainer}, {Ramos}, {Ramos-Lerate}, {Re Fiorentin}, {Regibo}, {Richards},
  {Rios Diaz}, {Ripepi}, {Riva}, {Rix}, {Rixon}, {Robichon}, {Robin}, {Robin},
  {Roelens}, {Rogues}, {Rohrbasser}, {Romero-G{\'o}mez}, {Rowell}, {Royer},
  {Ruz Mieres}, {Rybicki}, {Sadowski}, {S{\'a}ez N{\'u}{\~n}ez}, {Sagrist{\`a}
  Sell{\'e}s}, {Sahlmann}, {Salguero}, {Samaras}, {Sanchez Gimenez}, {Sanna},
  {Santove{\~n}a}, {Sarasso}, {Schultheis}, {Sciacca}, {Segol}, {Segovia},
  {S{\'e}gransan}, {Semeux}, {Shahaf}, {Siddiqui}, {Siebert}, {Siltala},
  {Silvelo}, {Slezak}, {Slezak}, {Smart}, {Snaith}, {Solano}, {Solitro},
  {Souami}, {Souchay}, {Spagna}, {Spina}, {Spoto}, {Steele},
  {Steidelm{\"u}ller}, {Stephenson}, {S{\"u}veges}, {Surdej}, {Szabados},
  {Szegedi-Elek}, {Taris}, {Taylor}, {Tolomei}, {Tonello}, {Torra}, {Torra},
  {Torralba Elipe}, {Trabucchi}, {Tsounis}, {Turon}, {Ulla}, {Unger},
  {Vaillant}, {van Dillen}, {van Reeven}, {Vanel}, {Vecchiato}, {Viala},
  {Vicente}, {Voutsinas}, {Weiler}, {Wevers}, {Wyrzykowski}, {Yoldas}, {Yvard},
  {Zhao}, {Zorec}, {Zucker}, \& {Zwitter}}]{GaiaCollab2022}
{Gaia Collaboration}, {Bailer-Jones}, C.~A.~L., {Teyssier}, D., {et~al.} 2022,
  arXiv e-prints, arXiv:2206.05681, \dodoi{10.48550/arXiv.2206.05681}

\bibitem[{{Graus} {et~al.}(2012){Graus}, {Lamb}, \& {Oey}}]{Graus2012}
{Graus}, A.~S., {Lamb}, J.~B., \& {Oey}, M.~S. 2012, \apj, 759, 10,
  \dodoi{10.1088/0004-637X/759/1/10}

\bibitem[{{Hillier} {et~al.}(2019){Hillier}, {Koenigsberger}, {Naz{\'e}},
  {Morrell}, {Barb{\'a}}, \& {Gamen}}]{Hillier2019}
{Hillier}, D.~J., {Koenigsberger}, G., {Naz{\'e}}, Y., {et~al.} 2019, \mnras,
  486, 725, \dodoi{10.1093/mnras/stz808}

\bibitem[{{Humphreys} \& {Davidson}(1994)}]{Humphreys1994}
{Humphreys}, R.~M., \& {Davidson}, K. 1994, \pasp, 106, 1025,
  \dodoi{10.1086/133478}

\bibitem[{{Humphreys} {et~al.}(2017{\natexlab{a}}){Humphreys}, {Davidson},
  {Hahn}, {Martin}, \& {Weis}}]{Humphreys2017a}
{Humphreys}, R.~M., {Davidson}, K., {Hahn}, D., {Martin}, J.~C., \& {Weis}, K.
  2017{\natexlab{a}}, \apj, 844, 40, \dodoi{10.3847/1538-4357/aa7cef}

\bibitem[{{Humphreys} {et~al.}(2017{\natexlab{b}}){Humphreys}, {Gordon},
  {Martin}, {Weis}, \& {Hahn}}]{Humphreys2017b}
{Humphreys}, R.~M., {Gordon}, M.~S., {Martin}, J.~C., {Weis}, K., \& {Hahn}, D.
  2017{\natexlab{b}}, \apj, 836, 64, \dodoi{10.3847/1538-4357/aa582e}

\bibitem[{{Humphreys} {et~al.}(2016){Humphreys}, {Weis}, {Davidson}, \&
  {Gordon}}]{Humphreys2016}
{Humphreys}, R.~M., {Weis}, K., {Davidson}, K., \& {Gordon}, M.~S. 2016, \apj,
  825, 64, \dodoi{10.3847/0004-637X/825/1/64}

\bibitem[{{Jones} {et~al.}(2015){Jones}, {Meixner}, {Sargent}, {Boyer},
  {Sewi{\l}o}, {Hony}, \& {Roman-Duval}}]{Jones2015}
{Jones}, O.~C., {Meixner}, M., {Sargent}, B.~A., {et~al.} 2015, \apj, 811, 145,
  \dodoi{10.1088/0004-637X/811/2/145}

\bibitem[{{Kalari} {et~al.}(2018){Kalari}, {Vink}, {Dufton}, \&
  {Fraser}}]{Kalari2018}
{Kalari}, V.~M., {Vink}, J.~S., {Dufton}, P.~L., \& {Fraser}, M. 2018, \aap,
  618, A17, \dodoi{10.1051/0004-6361/201833484}

\bibitem[{{Kalari} {et~al.}(2014){Kalari}, {Vink}, {Dufton}, {Evans},
  {Dunstall}, {Sana}, {Clark}, {Ellerbroek}, {de Koter}, {Lennon}, \&
  {Taylor}}]{Kalari2014}
{Kalari}, V.~M., {Vink}, J.~S., {Dufton}, P.~L., {et~al.} 2014, \aap, 564, L7,
  \dodoi{10.1051/0004-6361/201323289}

\bibitem[{{Koenigsberger} {et~al.}(2014){Koenigsberger}, {Morrell}, {Hillier},
  {Gamen}, {Schneider}, {Gonz{\'a}lez-Jim{\'e}nez}, {Langer}, \&
  {Barb{\'a}}}]{Koenigsberger2014}
{Koenigsberger}, G., {Morrell}, N., {Hillier}, D.~J., {et~al.} 2014, \aj, 148,
  62, \dodoi{10.1088/0004-6256/148/4/62}

\bibitem[{{Kraus}(2019)}]{Kraus2019}
{Kraus}, M. 2019, Galaxies, 7, 83, \dodoi{10.3390/galaxies7040083}

\bibitem[{{Lamb} {et~al.}(2016){Lamb}, {Oey}, {Segura-Cox}, {Graus}, {Kiminki},
  {Golden-Marx}, \& {Parker}}]{Lamb2016}
{Lamb}, J.~B., {Oey}, M.~S., {Segura-Cox}, D.~M., {et~al.} 2016, \apj, 817,
  113, \dodoi{10.3847/0004-637X/817/2/113}

\bibitem[{{Lamers} {et~al.}(1998){Lamers}, {Zickgraf}, {de Winter}, {Houziaux},
  \& {Zorec}}]{Lamers1998}
{Lamers}, H. J.~G.~L.~M., {Zickgraf}, F.-J., {de Winter}, D., {Houziaux}, L.,
  \& {Zorec}, J. 1998, \aap, 340, 117

\bibitem[{{Langer} {et~al.}(1994){Langer}, {Hamann}, {Lennon}, {Najarro},
  {Pauldrach}, \& {Puls}}]{Langer1994}
{Langer}, N., {Hamann}, W.~R., {Lennon}, M., {et~al.} 1994, \aap, 290, 819

\bibitem[{{Maeder} \& {Conti}(1994)}]{Maeder1994}
{Maeder}, A., \& {Conti}, P.~S. 1994, \araa, 32, 227,
  \dodoi{10.1146/annurev.astro.32.1.227}

\bibitem[{{Massey} {et~al.}(2014){Massey}, {Neugent}, {Morrell}, \&
  {Hillier}}]{Massey2014}
{Massey}, P., {Neugent}, K.~F., {Morrell}, N., \& {Hillier}, D.~J. 2014, \apj,
  788, 83, \dodoi{10.1088/0004-637X/788/1/83}

\bibitem[{{Massey} {et~al.}(2000){Massey}, {Waterhouse}, \&
  {DeGioia-Eastwood}}]{Massey2000}
{Massey}, P., {Waterhouse}, E., \& {DeGioia-Eastwood}, K. 2000, \aj, 119, 2214,
  \dodoi{10.1086/301345}

\bibitem[{{Oey} {et~al.}(2018){Oey}, {Dorigo Jones}, {Castro}, {Zivick},
  {Besla}, {Januszewski}, {Moe}, {Kallivayalil}, \& {Lennon}}]{Oey2018}
{Oey}, M.~S., {Dorigo Jones}, J., {Castro}, N., {et~al.} 2018, \apjl, 867, L8,
  \dodoi{10.3847/2041-8213/aae892}

\bibitem[{{Oksala} {et~al.}(2012){Oksala}, {Kraus}, {Arias}, {Borges
  Fernandes}, {Cidale}, {Muratore}, \& {Cur{\'e}}}]{Oksala2012}
{Oksala}, M.~E., {Kraus}, M., {Arias}, M.~L., {et~al.} 2012, \mnras, 426, L56,
  \dodoi{10.1111/j.1745-3933.2012.01323.x}

\bibitem[{{Pasquali} {et~al.}(2000){Pasquali}, {Nota}, {Langer},
  {Schulte-Ladbeck}, \& {Clampin}}]{Pasquali2000}
{Pasquali}, A., {Nota}, A., {Langer}, N., {Schulte-Ladbeck}, R.~E., \&
  {Clampin}, M. 2000, \aj, 119, 1352, \dodoi{10.1086/301257}

\bibitem[{{Phillips} {et~al.}(2024){Phillips}, {Oey}, {Cuevas}, {Castro}, \&
  {Kothari}}]{Phillips2024}
{Phillips}, G.~D., {Oey}, M.~S., {Cuevas}, M., {Castro}, N., \& {Kothari}, R.
  2024, \apj, 966, 243, \dodoi{10.3847/1538-4357/ad3909}

\bibitem[{{Renzo} {et~al.}(2019){Renzo}, {Zapartas}, {de Mink}, {G{\"o}tberg},
  {Justham}, {Farmer}, {Izzard}, {Toonen}, \& {Sana}}]{Renzo2019}
{Renzo}, M., {Zapartas}, E., {de Mink}, S.~E., {et~al.} 2019, \aap, 624, A66,
  \dodoi{10.1051/0004-6361/201833297}

\bibitem[{{Richardson} \& {Mehner}(2018)}]{Richardson2018}
{Richardson}, N.~D., \& {Mehner}, A. 2018, Research Notes of the American
  Astronomical Society, 2, 121, \dodoi{10.3847/2515-5172/aad1f3}

\bibitem[{{Smith}(2016)}]{Nathan2016}
{Smith}, N. 2016, \mnras, 461, 3353, \dodoi{10.1093/mnras/stw1533}

\bibitem[{{Smith}(2019)}]{Smith2019}
---. 2019, \mnras, 489, 4378, \dodoi{10.1093/mnras/stz2277}

\bibitem[{{Smith} \& {Owocki}(2006)}]{Smith2006}
{Smith}, N., \& {Owocki}, S.~P. 2006, \apjl, 645, L45, \dodoi{10.1086/506523}

\bibitem[{{Smith} \& {Tombleson}(2015)}]{Smith2015}
{Smith}, N., \& {Tombleson}, R. 2015, \mnras, 447, 598,
  \dodoi{10.1093/mnras/stu2430}

\bibitem[{{Solovyeva} {et~al.}(2020){Solovyeva}, {Vinokurov}, {Sarkisyan},
  {Atapin}, {Fabrika}, {Valeev}, {Kniazev}, {Sholukhova}, \&
  {Maslennikova}}]{Solovyeva2020}
{Solovyeva}, Y., {Vinokurov}, A., {Sarkisyan}, A., {et~al.} 2020, \mnras, 497,
  4834, \dodoi{10.1093/mnras/staa2117}

\bibitem[{{Virtanen} {et~al.}(2020){Virtanen}, {Gommers}, {Oliphant},
  {Haberland}, {Reddy}, {Cournapeau}, {Burovski}, {Peterson}, {Weckesser},
  {Bright}, {van der Walt}, {Brett}, {Wilson}, {Jarrod Millman}, {Mayorov},
  {Nelson}, {Jones}, {Kern}, {Larson}, {Carey}, {Polat}, {Feng}, {Moore}, {Vand
  erPlas}, {Laxalde}, {Perktold}, {Cimrman}, {Henriksen}, {Quintero}, {Harris},
  {Archibald}, {Ribeiro}, {Pedregosa}, {van Mulbregt}, \&
  {Contributors}}]{Scipy}
{Virtanen}, P., {Gommers}, R., {Oliphant}, T.~E., {et~al.} 2020, Nature
  Methods, 17, 261, \dodoi{https://doi.org/10.1038/s41592-019-0686-2}

\bibitem[{{Walborn} {et~al.}(2017){Walborn}, {Gamen}, {Morrell}, {Barb{\'a}},
  {Fern{\'a}ndez Laj{\'u}s}, \& {Angeloni}}]{Walborn2017}
{Walborn}, N.~R., {Gamen}, R.~C., {Morrell}, N.~I., {et~al.} 2017, \aj, 154,
  15, \dodoi{10.3847/1538-3881/aa6195}

\bibitem[{{Weis} \& {Bomans}(2020)}]{Weis2020}
{Weis}, K., \& {Bomans}, D.~J. 2020, Galaxies, 8, 20,
  \dodoi{10.3390/galaxies8010020}

\bibitem[{{Wisniewski} {et~al.}(2007){Wisniewski}, {Bjorkman}, {Bjorkman}, \&
  {Clampin}}]{Wisniewski2007}
{Wisniewski}, J.~P., {Bjorkman}, K.~S., {Bjorkman}, J.~E., \& {Clampin}, M.
  2007, \apj, 670, 1331, \dodoi{10.1086/522330}

\bibitem[{{Zaritsky} {et~al.}(2004){Zaritsky}, {Harris}, {Thompson}, \&
  {Grebel}}]{Zaritsky2004}
{Zaritsky}, D., {Harris}, J., {Thompson}, I.~B., \& {Grebel}, E.~K. 2004, \aj,
  128, 1606, \dodoi{10.1086/423910}

\bibitem[{{Zaritsky} {et~al.}(2002){Zaritsky}, {Harris}, {Thompson}, {Grebel},
  \& {Massey}}]{Zaritsky2002}
{Zaritsky}, D., {Harris}, J., {Thompson}, I.~B., {Grebel}, E.~K., \& {Massey},
  P. 2002, \aj, 123, 855, \dodoi{10.1086/338437}

\bibitem[{{Zickgraf} {et~al.}(1996){Zickgraf}, {Kovacs}, {Wolf}, {Stahl},
  {Kaufer}, \& {Appenzeller}}]{Zickgraf1996}
{Zickgraf}, F.~J., {Kovacs}, J., {Wolf}, B., {et~al.} 1996, \aap, 309, 505

\end{thebibliography}
\bibliographystyle{aasjournal}

\end{document}